\documentclass{aa}  

\usepackage{amsmath}
\usepackage{graphicx}
\usepackage{graphics}
\usepackage{txfonts}
\usepackage{natbib}
\usepackage{xcolor}

\bibpunct{(}{)}{;}{a}{}{,}

\def\teff{\ifmmode T_{\rm eff} \else $T_{\mathrm{eff}}$\fi}

\def\ltsima{$\buildrel<\over\sim$}
\def\lsim{\lower.5ex\hbox{\ltsima}}

\newcommand{\hii}{H~{\sc ii}}
\newcommand{\ha}{\ifmmode {\rm H}\alpha \else H$\alpha$\fi}
\newcommand{\hb}{\ifmmode {\rm H}\beta \else H$\beta$\fi}

\newcommand{\heii}{He~{\sc ii}}

\newcommand{\nv}{N~{\sc v}}
\newcommand{\Heiiuv}{He~{\sc ii} $\lambda$1640}

\newcommand{\ebv}{\ifmmode E_{\rm B-V} \else $E_{\rm B-V}$\fi}
\newcommand{\av}{\ifmmode A_{\rm V} \else $A_{\rm V}$\fi}

\def\msun{\ifmmode M_{\odot} \else M$_{\odot}$\fi}
\def\msunyr{\ifmmode M_{\odot} {\rm yr}^{-1} \else M$_{\odot}$ yr$^{-1}$\fi}
\def\zsun{\ifmmode Z_{\odot} \else Z$_{\odot}$\fi}

\def\lsun{\ifmmode L_{\odot} \else L$_{\odot}$\fi}

\def\mup{\ifmmode M_{\rm up} \else M$_{\rm up}$\fi}
\def\mlow{\ifmmode M_{\rm low} \else M$_{\rm low}$\fi}

\def\aap{A\&A}

\def\apjl{ApJL}
\def\apjs{ApJS}

\newcommand{\oh}{\ifmmode 12 + \log({\rm O/H}) \else$12 + \log({\rm
O/H})$\fi}

\newcommand{\Nvuv}{N~{\sc v} $\lambda$1240}
\newcommand{\Nivuvfar}{N~{\sc iv} $\lambda$1719}
\newcommand{\Nivuvnear}{N~{\sc iv} $\lambda$1486}

\def\flyf{\ifmmode f_{\rm Lyf} \else $f_{\rm Lyf}$\fi}
\def\pz{\ifmmode P(z) \else $P(z)$\fi}
\def\ki2{\ifmmode \chi^2 \else $\chi^2$\fi}
\def\zphot{\ifmmode z_{\rm phot} \else $z_{\rm phot}$\fi}

\newcommand{\xphot}{\ifmmode x_\gamma \else $v_\gamma$\fi}
\newcommand{\xobs}{\ifmmode x_{\rm obs} \else $x_{\rm obs}$\fi}
\newcommand{\xcmf}{\ifmmode x_{\rm CMF} \else $x_{\rm CMF}$\fi}
\newcommand{\vexp}{\ifmmode V_{\rm exp} \else $V_{\rm exp}$\fi}
\newcommand{\vmax}{\ifmmode V_{\rm max} \else $V_{\rm max}$\fi}
\newcommand{\nh}{\ifmmode N_{\rm HI} \else $N_{\rm HI}$\fi}
\newcommand{\dv}{\ifmmode \Delta v({\rm em-abs}) \else $\Delta v({\rm em}-{\rm abs})$\fi}

\def\fesc{\ifmmode f_{\rm esc} \else $f_{\rm esc}$\fi}
\def\fescrel{\ifmmode f_{\rm esc,rel} \else $f_{\rm esc,rel}$\fi}

\def\frellya{\ifmmode f^{\rm rel}_{\rm{Ly}\alpha} \else $f^{\rm rel}_{\rm{Ly}\alpha}$\fi}

\newcommand{\mstar}{\ifmmode M_\star \else $M_\star$\fi}
\newcommand{\muv}{\ifmmode M_{1500} \else $M_{1500}$\fi}
\newcommand{\auv}{\ifmmode A_{\rm UV} \else $A_{\rm UV}$\fi}
\newcommand{\luv}{\ifmmode L_{\rm UV} \else $L_{\rm UV}$\fi}
\newcommand{\lir}{\ifmmode L_{\rm IR} \else $L_{\rm IR}$\fi}
\newcommand{\lbol}{\ifmmode L_{\rm bol} \else $L_{\rm bol}$\fi}
\newcommand{\liruv}{\ifmmode L_{\rm IR+UV} \else $L_{\rm IR+UV}$\fi}
\newcommand{\liroveruv}{\ifmmode L_{\rm IR}/L_{\rm UV} \else $L_{\rm IR}/L_{\rm UV}$\fi}
\newcommand{\nlyc}{\ifmmode N_{\rm Lyc} \else $N_{\rm Lyc} $\fi}
\newcommand{\rholyc}{\ifmmode \rho_{\rm Lyc} \else $\rho_{\rm Lyc} $\fi}
\newcommand{\chion}{\ifmmode \xi_{\rm ion} \else $\xi_{\rm ion}$\fi}
\newcommand{\chioncorr}{\ifmmode \xi_{\rm ion}^0 \else $\xi_{\rm ion}^0$\fi}

\newcommand{\Civuv}{C~{\sc iv} $\lambda$1550}
\newcommand{\Civ}{C~{\sc iv}}

\usepackage{hyperref}  
\hypersetup{colorlinks=true,linkcolor=[rgb]{1.,0.2,0.2},citecolor=[rgb]{0.1,0.4,1.},filecolor=[rgb]{0.7,0.2,0.2},urlcolor=[rgb]{0.1,0.5,0.2}}

\begin{document}

\title{Evidence for very massive stars in extremely UV-bright star-forming galaxies at $z \sim 2.2 - 3.6$}
\subtitle{}
\authorrunning{A. Upadhyaya et al.}
\author{A. Upadhyaya\inst{\ref{unige},\ref{warwick}},
R. Marques-Chaves\inst{\ref{unige}},
D. Schaerer\inst{\ref{unige},\ref{CNRS}} 
F. Martins\inst{\ref{lupm}}, 
I. P\'{e}rez-Fournon\inst{\ref{iac},\ref{ull}},
A. Palacios\inst{\ref{lupm}}, 
E. R. Stanway\inst{\ref{warwick}}
}

\institute{Geneva Observatory, Department of Astronomy, University of Geneva, Chemin Pegasi 51, CH-1290 Versoix, Switzerland \label{unige}
\and
Department of Physics, University of Warwick, Gibbet Hill Road, Coventry, CV4 7AL, UK \label{warwick}
\and
CNRS, IRAP, 14 Avenue E. Belin, 31400 Toulouse, France \label{CNRS}
\and
LUPM, Universit\'e de Montpellier, CNRS, Place Eug\`ene Bataillon, F-34095 Montpellier, France
\label{lupm}
\and
Instituto de Astrof\'\i sica de Canarias, C/V\'\i a L\'actea, s/n, E-38205 San Crist\'obal de La Laguna, Tenerife, Spain \label{iac}
\and
Universidad de La Laguna, Dpto. Astrof\'\i sica, E-38206 San Crist\'obal de La Laguna, Tenerife, Spain \label{ull}
}

\titlerunning{VMS in UV-bright galaxies}

\date{Received date; accepted date}

\abstract{We present a comprehensive analysis of the presence of very massive stars (VMS > $100 M_{\odot}$) in the integrated spectra of 13 UV-bright star-forming galaxies at $2.2 \lesssim z \lesssim 3.6$ taken with the Gran Telescopio Canarias (GTC). These galaxies have very high UV absolute magnitudes ($M_{\rm UV} \simeq -24$), intense star formation (star formation rate $ \simeq 100-1000$ $M_{\odot}$ yr$^{-1}$), and metallicities in the range of 12+log(O/H) $\simeq8.10-8.50$ inferred from strong rest-optical lines. The GTC rest-UV spectra reveal spectral features indicative of very young stellar populations with VMS, such as strong P-Cygni line profiles in the wind lines N~{\sc v} $\lambda 1240$ and C~{\sc iv} $\lambda 1550$ along with intense and broad He~{\sc ii} $\lambda 1640$ emission with equivalent width ($EW_{0}$)  $\simeq 1.40-4.60$ \AA, and full width half maximum (FWHM) $\simeq 1150-3170$ $km \ s^{-1}$. A Comparison with known VMS-dominated sources and typical galaxies without VMS reveals that some UV-bright galaxies closely resemble VMS-dominated clusters (e.g., R136 cluster). The presence of VMS is further supported by a quantitative comparison of the observed strength of the He~{\sc ii} emission with population synthesis models with and without VMS, where models with VMS are clearly preferred. Employing an empirical threshold for $EW_{0}$ (\heii) $\geq 3.0$ \AA, along with the detection of other VMS-related spectral profiles (N~{\sc iv} $\lambda 1486, 1719$), we classify nine out of 13 UV-bright galaxies as VMS-dominated sources. This high incidence of VMS-dominated sources in the UV-bright galaxy population ($\approx 70\%$) contrasts significantly with the negligible presence of VMS in typical $L_{\rm UV}^{*}$ LBGs at similar redshifts ($<1\%$). Our results thus indicate that VMS are common in UV-bright galaxies, suggesting a different initial mass function (IMF) with upper mass limits between $175 M_{\odot}$ and $475 M_{\odot}$.}

 \keywords{Galaxies: starburst -- Galaxies: high-redshift -- Ultraviolet: galaxies -- Stars: massive}

 \maketitle


\section{Introduction}
\label{s_intro}

The epoch of cosmic noon at a redshift around two to three plays an important role in the evolution of our Universe. It is the time when galaxies underwent intense star formation with a peak in the star formation rate density \citep{lilly1996, feulner2005, madau2014, lopezfernandez2018, sanchez2019, Koushan2021}. 

Typical star-forming galaxies at cosmic noon show strong Ly$\alpha$~$\lambda 1216$ emission along with stellar wind features such as \Nvuv \ and \Civuv \ P-Cygni line profiles in their rest-frame UV 
spectra \cite[e.g.,][]{shapley2003, berry2012, steidel2016, lefevre2019, marques2020}. 
Due to their faintness, with UV absolute magnitudes of $M_{UV} \gtrsim -21$, the detailed characterization of their rest-frame UV has been achieved using stacking techniques of hundreds to thousands of individual spectra \citep[e.g.,][]{shapley2003, lefevre2015, steidel2016}, very deep spectroscopy \cite[e.g.,][]{Pentericci2018, Garilli2021}, or with the help of gravitational lensing \citep[e.g.,][]{pettini2000, quider2009, mirka2010, marques2017, marques2020}. Even with many extensive surveys before the launch of JWST, only a small number of star-forming galaxies brighter than $M_{UV} \leq -22$ were discovered at this cosmic epoch \citep[e.g.,][]{bian2012}. 

Recently, \citet{marques2020b, marques2021, marques2022} found star-forming galaxies with $M_{UV} \sim -24$.
Their brightness and very small number density suggest an extreme star-formation rate (SFR) which consumes the gas rapidly. Additionally, these UV bright galaxies show intense and broad \Heiiuv \ emission with rest-frame equivalent widths $(EW_{0}) \gtrsim 3 \AA $ and full width half maximum (FWHM) of $\simeq 1000-3000$ km s$^{-1}$, which is not common for star-forming galaxies at this redshift. Last, but not least, some of these galaxies have been found to be strong emitters of Lyman continuum radiation, posing new questions on the main contributors of cosmic reionization \citep[see,][]{marques2021}.

Determining the stellar content of these UV-bright star-forming galaxies is crucial to understanding the origin and evolution of these sources. The broad \Heiiuv \ emission provides some clues on the stellar populations that dominate the rest-frame UV spectra of these galaxies. The R136 cluster in the Large Magellanic Cloud (LMC) shows a similar broad \Heiiuv \ profile \citep[$EW_{0} \sim 4.5 \AA $, FWHM $\simeq 1800$ km s$^{-1}$;][]{crowther2010, crowther2016}. The R136 cluster has been confirmed to host very massive stars (VMS > 100 $M_{\odot}$) \citep{crowther2010, bestenlehner2011, Hainich2014, crowther2016, bestenlehner2020, brands2022} with the most recent studies showing its most massive star to have a mass of at least 200 $M_{\odot}$ \citep{Kalari2022,shenar23}. 

There have been other studies in the local Universe where the presence or potential presence of VMS has been detected \citep[e.g.,][]{Massey1998, Crowther1998, Bruhweiler2003, Martins2008}. Using Hubble Space Telescope (HST) UV spectroscopy, \cite{Wofford2014} studied the super star cluster A1 in NGC3125 and inferred the presence of VMS; a further study strengthened this interpretation \citep{Wofford2023}. \cite{Smith2016} inferred the presence of VMS in one of the nuclear star clusters in the compact \hii \ region of the blue dwarf galaxy NGC 5253 using HST UV and Very Large Telescope (VLT) optical spectroscopy. \cite{senchyna2021} analyzed HST UV spectra of a few nearby star-forming regions showing Wolf-Rayet (WR) features in the Sloan Digital Sky Survey (SDSS) spectrum and argued that VMS may be present in some of them. \cite{Smith2023} found evidence for VMS in a super star cluster in the metal-poor galaxy Mrk 71. Recently, \cite{Mestric2023} noted the presence of VMS in a gravitationally lensed star cluster, the Sunburst cluster at $z=2.37$. The recent study of \cite{Martins2023} not only infers the presence of VMS in a few nearby star-forming regions but also provides guidance on how to separate VMS from WR sources using both UV and optical spectroscopy. All of these studies suggest that sources hosting VMS have intense and broad \Heiiuv\ emission with a relatively high $EW_{0}$, typically EW (\heii) $\ga 2-3$ \AA \ and FWHM $\ga$ 1000 km s$^{-1}$.

So far, little is known about the presence and occurrence of VMS in high-redshift galaxies. While broad \Heiiuv\ emission is a fairly common feature of distant star-forming galaxies, it is relatively weak ($EW_{0}$ $\sim 1-2 \ \AA$, FWHM $\ga$ 1000 km s$^{-1}$) and attributed to WR stars \citep[see e.g.,][]{shapley2003}. Stronger and broad \heii\ emission seems rare and has been noted for example \ in studies by \cite{cassata2013}, \cite{Nanayakkara2019}, \cite{saxena2020}, and \cite{Wofford2023} in Lyman break galaxies (LBGs).

Here, we report the detection of VMS in a large fraction of extremely UV-bright star-forming galaxies at z $\sim$ 2.2 - 3.6. Using observations taken with the Optical System for Imaging and low-Intermediate-Resolution Integrated Spectroscopy (OSIRIS) spectrograph at the Gran Telescopio Canarias (GTC), and the latest VMS models from \cite{fabrice2022}, we present a detailed analysis of the rest-UV spectra and the main spectral features of 13 UV-bright galaxies discovered by \citet{marques2020, marques2021, marques2022}. These objects consistently show very strong \Heiiuv\ emission and other stellar features, providing unique insight on their stellar populations and the high mass end of the initial mass function.

The paper is structured as follows. In Section \ref{s_sample}, we describe our observations and how the data have been reduced. In Section \ref{s_sources}, we describe the properties of the UV bright galaxies and present their \Heiiuv \ emission spectral profiles. In Section \ref{ss_empirical_results}, we provide empirical evidence for the presence of VMS in our sources. In Section \ref{ss_models_results}, we present the models and their analysis to complement the evidence for the presence of VMS in our sources. In Section \ref{s_discussion}, we discuss various implications of these results. Finally, Section \ref{s_conclusion} provides us with conclusions and summarizes the results of our work. Throughout this work,we assume a concordance cosmology with $\Omega_{m}$ = 0.274, $\Omega_{\Lambda}$ = 0.726, and $H_{0}$ = 70 km s$^{-1}$ Mpc$^{-1}$. All magnitudes are given in the AB system.


\section{Observations}
\label{s_sample}

\subsection{Sample selection}

The galaxies studied in this work are part of a large sample of $\sim 70$ UV-luminous star-forming galaxies at $z \simeq 2.0-3.6$ selected from the $\sim 9000$ deg$^{2}$-wide extended Baryon Oscillation Spectroscopic Survey (eBOSS; \citealt{abolfathi2018}) of the SDSS \citep[][]{eisenstein2011}. The sample, selection techniques, and overall properties will be presented in a separate work (R.~Marques-Chaves in prep.). Here, we explore the properties of 13 of these UV-bright sources that have deep follow-up spectroscopy. The properties of three of these sources were already analyzed in detail in \cite{marques2020b, marques2021, marques2022} and \cite{alvarez2021}, while the remaining are new discoveries. 

Table \ref{tab1} shows several properties of these 13 sources, including coordinates, redshifts, and magnitudes. Overall, these sources are optically bright, $R \simeq 20.8 - 21.9$, corresponding to rest-frame UV absolute magnitudes from $M_{\rm UV} = -23.3$ to $-24.6$. Given their brightness, they are ideal targets for deep, high signal-to-noise (S/N) spectroscopy.

\begin{table*}
\begin{center}
\caption{Summary of the GTC observations. \label{tab1}}
\resizebox{1.0\textwidth}{!}{%
\begin{tabular}{l c c c c c c c c}
\hline \hline
\smallskip
\smallskip
Name & R.A. &  Dec. & $z$ & $R$-band &  \multicolumn{2}{c}{Optical spectroscopy} & \multicolumn{2}{c}{Near-IR spectroscopy} \\
\cline{6-9}
  &  &  &   &     & Date & Exp. time & Date & Exp. time \\
  & (J2000) & (J2000) &   &  (AB) &  (dd/mm/yyyy) & (sec) & (dd/mm/yyyy) & (sec) \\
\hline 
  J0006+2452 &  00:06:44.73  & +24:52:53.19  & 2.379  & 20.98  & 16/07/2020 & 3600 & 12/07/2019 & 2560 \\
  J0031+3545 &  00:31:12.43  & +35:45:56.12  & 2.816  & 20.84  & 16/07/2020 & 3600 & --- & --- \\
  J0036+2725 &  00:36:06.10  & +27:25:39.27  & 2.171  & 21.07  & 19/08/2020 & 3600 & --- & --- \\
  J0110$-$0501 & 01:10:45.58 & $-$05:01:39.27 & 2.368 & 21.80 & 30/08/2019 & 3720 & 20/09/2018 & 2560 \\
  J0115+1837   &  01:15:21.95 & +18:37:44.47  & 2.322  & 21.59  & 23/09/2019 & 3720 & 20/09/2018 & 2560 \\
  J0121+0025$^{a}$ &  01:21:56.09  & +00:25:20.30  & 3.246  & 21.60 & 18/08/2020 & 7200 & --- & --- \\
  J0146$-$0220 & 01:46:37.02 & $-$02:20:55.86 & 2.160  & 21.28 & 10/08/2018 & 4500 & 19/09/2018 & 2560 \\ 
  J0850+1549 &  08:50:38.86  & +15:49:17.88  & 2.424  & 21.43 & 12/03/2018 & 3600 & 25/04/2018 & 2560\\
  J1013+4650 &  10:13:28.76  & +46:50:43.47  & 2.286  & 21.40 & 07/11/2018 & 4500 & 26/12/2018 & 2560 \\
  J1157+0113 &  11:57:34.12  & +01:13:08.21  & 2.545  & 21.88 & 01/05/2019 & 7200 & --- & --- \\
  J1220+0842$^{b}$ & 12:20:40.72 & +08:42:38.14 & 2.470 & 20.86 & 19/06/2017 & 3000 & 23/03/2018 & 2560 \\
  J1316+2614$^{c}$ & 13:16:29.61 & +26:14:07.05 & 3.612 & 21.26 & 20/04/2021 & 9000 & 08/05/2018 &  2560\\
  J1335+4330 &  13:35:15.79  & +43:30:35.07  & 2.170  & 21.34 & 11/04/2018 & 3600 & 25/04/2018 & 2560 \\

\hline 
\end{tabular}}
\end{center}
\textbf{Notes. ---} data already presented in a) \cite{marques2021}, b) \cite{marques2020b}, and c) \cite{marques2022}.
\end{table*}

\subsection{Optical spectroscopy}

Optical spectra were obtained with the OSIRIS\footnote{\url{http://www.gtc.iac.es/instruments/osiris/}} instrument on the GTC between 2017 and 2021 (Table \ref{tab1}) under the GTC programs IDs: GTC67-17A, GTCMULTIPLE2F-18A, GTC50-18A, GTCMULTIPLE2E-18B, GTCMULTIPLE2E-19A, GTC21-20A, and GTC29-21A (PI: R. Marques-Chaves).
Observations were performed under good seeing conditions ($\lesssim 1.2^{\prime \prime}$, FWHM) using the low-resolution grism R1000B, which provides a spectral resolution $\rm R\sim 700$ and coverage of 3600-7600\AA. Long-slits with 1.2$^{\prime \prime}$-width were centered on each object and oriented with the parallactic angle. Total on-source exposure times vary from 3000 to 9000 seconds, depending on the source (Table \ref{tab1}), and were split into at least four individual exposures for cosmic rays removal. 

Data were reduced following standard reduction procedures using {\sc iraf}, starting from the subtraction of the bias and flat-field correction. The wavelength calibration is done using HgAr+Ne+Xe arc-lamp data. 2D spectra are background subtracted using sky regions on both sides of the trace of each source. Individual 1D spectra are extracted, stacked, and corrected for the instrumental response using observations of standard stars observed each night. We use the extinction curve of \cite{cardelli1989} and the extinction map of \cite{schlafly2011} to correct for the reddening effect in the Galaxy. Finally, we corrected for telluric absorption using the {\sc iraf} \textit{telluric} routine. Individual 1D spectra are shown in Appendix \ref{A1_GTC_spectra} and Figure \ref{spectra_source}.

\subsection{Near-IR spectroscopy}

We also obtained near-IR spectra for nine sources with the Espectr\'ografo Multiobjeto Infra-Rojo (EMIR)\footnote{\url{http://www.gtc.iac.es/instruments/emir/}} on the GTC between 2018 and 2019. For each source, we use the $HK$ grism with a $0.8^{\prime \prime}$-width, providing a spectral resolution $\rm R\sim 700$ and coverage of 1.45-2.41$\mu$m. The slits were centered using a bright reference star. Observations were taken with on-source exposure times of 16$\times$160 sec. with a standard 10$^{\prime \prime}$ ABBA dither. Reduction of near-IR spectra was performed using the official EMIR pipeline\footnote{\url{https://pyemir.readthedocs.io/en/latest/index.html}}. 

\section{Properties of UV-bright galaxies}
\label{s_sources}

\begin{figure*}[h]
    \centering
    \includegraphics[width=\linewidth]{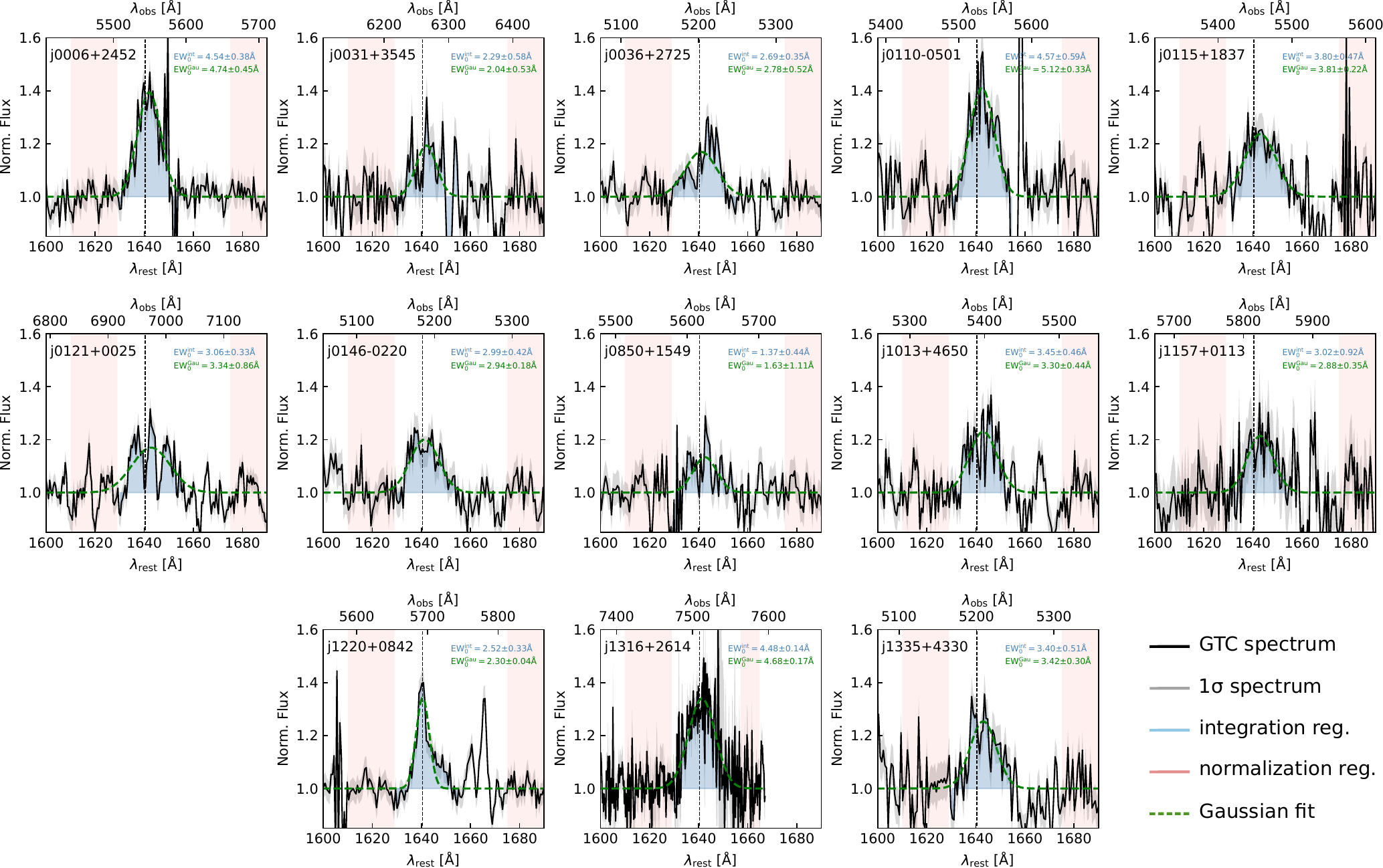}
    \caption{Normalized GTC spectra of the 13 UV-bright galaxies studied in this work around the \Heiiuv \ line (black, and 1$\sigma$ uncertainty in gray). Spectra are normalized using the continuum regions marked in red. Equivalent widths of the He~{\sc ii} line are measured by integrating the spectral regions from 1630 \AA \ to 1655 \AA \ (blue regions). Green dashed lines show Gaussian fits of the \Heiiuv \ emission. The x-axes represent the rest-frame and observed wavelengths (bottom and top axis, respectively). }
    \label{HeII_profiles}
\end{figure*}

From the results of Marques-Chaves et al. (in prep.) the UV-bright galaxies show very steep UV slopes ($f_{\lambda} \propto \lambda^{\beta}$) between $\beta \simeq -1.7$ and $\beta \simeq -2.8$ with a mean and scatter of $<\beta > = -2.33 \pm 0.30$. This translates into little dust obscuration, with a mean value of $E(B-V) = 0.05 \pm 0.04$ for this sample assuming the \cite{calzetti2000} extinction law with an intrinsic $\beta_{0} = -2.44$. SFRs are derived from the UV-luminosity, using the specific conversion factor $\kappa_{\rm UV} = 1.3 \times 10^{-28} M_{\odot}$~yr$^{-1}$/(erg s$^{-1}$ Hz$^{-1}$) derived using Binary Population and Spectral Synthesis (BPASS) binary models \citep{bpass2017, bpass2018, bpass2022} that assumes the \cite{chabrier2003} initial mass function (IMF) with an upper mass cutoff of 100 $M_{\odot}$, the LMC metallicity, and a continuous star-formation over 10~Myr (the typical age observed for these sources)\footnote{Note that this value of $\kappa_{\rm UV}$ implies SFR values higher by a factor $\sim 1.7$ than the classical assumption of SFR=const over $\sim 100$ Myr.}. We correct the derived UV-SFRs for the dust attenuation using the observed UV slopes and assuming the \cite{calzetti2000} attenuation law. The dust-corrected SFRs of our sources range over $\simeq 100-900$~$M_{\odot}$~yr$^{-1}$ and are listed in Table \ref{table_measurements}. The stellar masses of the young stellar population are derived assuming a continuous SFR over 10 Myr, and are also listed in Table \ref{table_measurements}. 
We refer to R. Marques-Chaves et al. (in prep.) for the description of the general properties of this sample. The properties of a few of these sources, namely J1220+0842, J0121+0025, and J1316+2614, were already investigated in detail in \cite{marques2020b, marques2021, marques2022}, for which they find very young stellar populations ($\simeq 10$~Myr) with $\rm SFR \simeq 400-600 M_{\odot}$ yr$^{-1}$ and stellar masses of log($M_{\star}/M_{\odot}) \simeq 9.6-9.8$. J0121+0025 and J1316+2614 are also found to be strong Lyman continuum (LyC) emitters, with LyC escape fractions of $\approx 40\% - 90\%$ \citep{marques2021, marques2022}.

\begin{table*}
\begin{center}
\caption{Global properties and measurements of the strengths of stellar \Heiiuv \ emission, \Nvuv, and \Civuv \ absorption. The \Heiiuv \ emission $EW_{0}$ is measured by integrating normalized flux in the spectral regions from 1630 \AA \ to 1655 \AA. The \Nvuv \ and \Civuv \ absorption $EW_{0}$ are measured by integrating normalized flux in the spectral regions from 1230 \AA \ to 1240 \AA, and 1530 \AA \ to 1543 \AA \ respectively}. \label{table_measurements}.
\begin{tabular}{lcccccccc}
\hline \hline
\smallskip
\smallskip
Name &   $M_{\rm UV}$   &  $SFR$  & $M_{\star}^{\rm young}$ & 12+log(O/H) & $EW_{0}$ (N~{\sc v}) & $EW_{0}$ (C~{\sc iv}) & $EW_{0}$ (He~{\sc ii})  & FWHM (He~{\sc ii})  \\
 &  (AB) &  ($M_{\odot}$~yr$^{-1}$) & (log[$M_{\odot}$]) &  & (\AA) &  (\AA) & (\AA) & (km s$^{-1}$) \\
\hline
J0006+2452  &  $-24.17$ & $316 \pm 31$ & $9.50 \pm 0.05$ & $8.49 \pm 0.10$ &   $-3.78\pm 0.40$ & $-6.70\pm 0.19$ & $4.54 \pm 0.38$ & $2070\pm180$  \\ 
J0031+3545  &  $-24.62$ & $390 \pm 60$ & $9.59 \pm 0.07$ & --- & $-2.11 \pm 0.48$ & $-5.40\pm 0.34$ & $2.29 \pm 0.58$ & $1800 \pm 430$ \\
J0036+2725 &   $-23.91$ & $612 \pm 62$ & $9.79 \pm 0.05$ & --- & $-5.40 \pm 0.34$ & $-6.29\pm 0.19$  & $2.69 \pm 0.35$ & $2810 \pm 470$   \\ 
J0110$-$0501 & $-23.33$ & $110 \pm 15$ & $9.04 \pm 0.06$ & $8.48 \pm 0.15$ & $-4.87 \pm 0.64$ & $-4.32\pm 0.35$ & $4.57 \pm 0.59$ & $1990 \pm 510$ \\
J0115+1837 &   $-23.52$ & $210 \pm 35$ & $9.33 \pm 0.07$ & $8.37 \pm 0.12$ & $-2.08 \pm 0.76$ & $-4.22 \pm 0.29$ & $3.80 \pm 0.47$ & $2870 \pm 165$  \\ 
J0121+0025 &   $-24.11$ & $460 \pm 90$ & $9.65 \pm 0.08$ & --- & $-3.54 \pm 0.20$ & $-5.51 \pm 0.20$ & $3.06 \pm 0.33$ & $3170 \pm 580$  \\ 
J0146$-$0220&  $-23.68$ & $340 \pm 43$ & $9.53 \pm 0.06$ & $8.47 \pm 0.20$ & $-3.38 \pm 0.46$ & $-3.87 \pm 0.27$ & $2.99 \pm 0.42$ & $2480 \pm 135$ \\ 
J0850+1549  &  $-23.76$ & $190 \pm 26$ & $9.28 \pm 0.06$ & $8.27 \pm 0.13$ & $-1.18 \pm 0.46$ & $-2.27 \pm 0.31$ & $1.37 \pm 0.44$ & $2030 \pm 1351$ \\ 
J1013+4650  &  $-23.67$ & $210 \pm 27$ & $9.32 \pm 0.05$ & $8.45 \pm 0.10$ & $-2.64 \pm 0.53$ & $-6.16 \pm 0.27$ & $3.45 \pm 0.46$ & $2520 \pm 310$  \\ 
J1157+0113  &  $-23.40$ & $195 \pm 68$ & $9.29 \pm 0.13$ & --- & $-2.32 \pm 1.12$ & $-2.41 \pm 0.62$ & $3.02 \pm 0.92$ & $2280 \pm 260$  \\ 
J1220+0842  &  $-24.36$ & $360 \pm 32$ & $9.56 \pm 0.04$ & $8.13 \pm 0.19$ & $-3.16 \pm 0.29$ & $-2.00 \pm 0.21$ & $2.52 \pm 0.33$ & $1150 \pm 20$  \\ 
J1316+2614  &  $-24.65$ & $415 \pm 115$ & $9.62 \pm 0.11$ & $8.45 \pm 0.12$ & $-1.87 \pm 0.68$ & $-3.82 \pm 0.54$ & $4.48 \pm 0.14$ & $2370 \pm 85$  \\ 
J1335+4330  &  $-23.64$ & $920 \pm 150$ & $9.96 \pm 0.08$ & $8.33 \pm 0.17$ & $-2.70 \pm 0.88$ & $-3.64 \pm 0.34$ & $3.40 \pm 0.51$ & $2350 \pm 200$   \\ 

\hline 
\end{tabular}
\end{center}
\end{table*}

\subsection{Strong He~{\sc ii} $\lambda1640$ and other wind lines}\label{sec31}

The GTC spectra present a high S/N in the continuum, revealing a wealth of spectral features arising from different galaxy components. These include ISM absorption lines (e.g., Si~{\sc ii} $\lambda 1260$, C~{\sc ii} $\lambda 1334$), photospheric absorption lines (e.g., S~{\sc v} $\lambda 1501$), or nebular gas for some sources (O~{\sc iii}] $\lambda 1666$, C~{\sc iii}] $\lambda 1908$). 

In particular, and most relevant for the present work, is the detection of intense emission in He~{\sc ii} $\lambda 1640$ for several sources in our sample. Figure \ref{HeII_profiles} shows the He~{\sc ii} profiles observed in the spectra of our sources. 
We measure the strength of \heii \ emission in our sources by measuring the rest-frame equivalent width ($EW_{0}$) using an integration region from 1630~\AA \ to 1655~\AA{ }(rest; blue in Figure \ref{HeII_profiles}), and two pseudo-continuum regions on both sides of \heii \ for the continuum estimation ($1610-1629$~\AA \ and $1675-1690$~\AA, respectively; red in Figure \ref{HeII_profiles}). The continuum regions were selected to avoid the contribution of the low-ionization ISM lines Fe~{\sc ii} $\lambda 1608$ and Al~{\sc ii} $\lambda 1670$ and the nebular emission from [O~{\sc iii}] $\lambda \lambda 1661,1666$. We measure $EW_{0}$(\heii) between $\simeq 1.4$ \AA \ and $\simeq 4.6$ \AA \ for our sources (see Table \ref{table_measurements}). In particular, J0006+2452, J0110-0501, and J1316+2614 are the most extreme \heii \ emitters in our sample showing $EW_{0}$(\heii) $\simeq 4.5$ \AA, which is similar to that measured in local star-forming regions with VMS (e.g., R136 star cluster with $EW_{0}$(\heii)$ \simeq 4.5$ \AA; \citealt{crowther2016}). 

We also fit Gaussian profiles to the continuum-normalized He~{\sc ii} emission. The best-fits are shown in green in Figure \ref{HeII_profiles}. The He~{\sc ii} emission appears broad for all sources with $\rm FWHM \simeq 1000-3000$ km s$^{-1}$. The He~{\sc ii} line appears symmetric for some sources (e.g., J0006+2452 or J0115+1837), while for others the line appears asymmetric (J1220+0842 or J1335+4330) or with multiple emission/absorption peaks (e.g., J0031+3545 or J0121+0025).

The GTC spectra also reveal strong P-Cygni line profiles in the wind lines N~{\sc v} $\lambda 1240$ and C~{\sc iv} $\lambda 1550$, but also N~{\sc iv} $\lambda 1486$, and N~{\sc iv} $\lambda 1719$ for some sources (see Appendix \ref{A1_GTC_spectra}). These profiles are the result of strong stellar winds from O-type stars and indicate very young ages ($\lesssim 10$~Myr) of the stellar population \citep[e.g.,][]{leitherer2011}. Well-defined P-Cygni line profiles in Si~{\sc iv} $\lambda 1400$, which are characteristic of O-type supergiants \citep[e.g.,][]{walborn85,garcia2004}, are also detected in some sources (e.g., J0006$+$2452, J0110$-$0501, J1316$+$2614).

\subsection{Metallicity}

An important property worth discussing is the metallicity of these sources, which governs the mass-loss rate of massive stars and thus the shape and strength of wind lines. Strong P-Cygni profiles are detected in the GTC spectra and, in particular, in C~{\sc iv} which is known to be the most sensitive feature to the metallicity \citep[][]{chisholm2019}. This suggests that the metal content in these sources cannot be extremely low (hence $Z/Z_{\odot} \gtrsim 0.1$). However, an accurate determination of the stellar metallicity using wind lines is rather difficult given the natural degeneracy between metallicity, age, and the shape of the IMF (slope and upper mass limit). Similarly, metallicity indicators using the strength of photospheric absorption features \citep[e.g.,][]{rix2004, sommariva2012, calabro2021} may not work properly for these sources as well since they were calibrated using stellar models assuming standard IMF and ages of 100~Myr. 

\begin{figure}
    \centering
    \includegraphics[width=\linewidth]{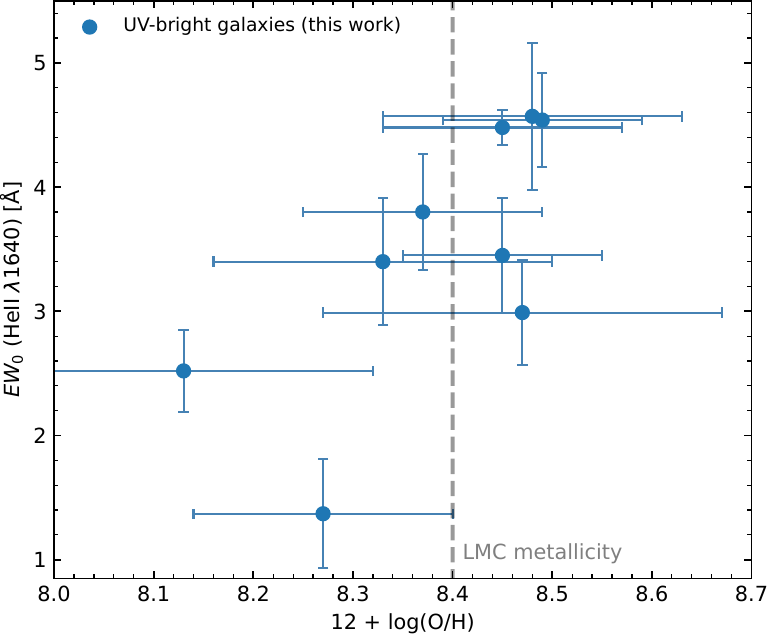}
    \caption{Relationship between the strength of He~{\sc ii} $\lambda 1640$ ($EW$ in \AA) and the oxygen abundance of our sources. The LMC metallicity is marked with the vertical dashed line.}
    \label{fig_met}
\end{figure}

To overcome this, we perform a sanity check on the nebular metallicity using rest-frame optical lines. Specifically, we extract the flux of H$\alpha$ and [N~{\sc ii}] $\lambda 6585$ emission lines using Gaussian profiles and relate the observed $N2 = \log$ ([N~{\sc ii}]$/\rm H\alpha$) line ratio with metallicity using the strong-line calibrator of \cite{marino2013}. For the seven sources with observations of H$\alpha$ and [N~{\sc ii}], we measure values of $N2$ between $-1.02$ to $-0.55$, yielding 12+log(O/H) $= 8.27-8.49$. We also used the metallicity measurements of J1220+0842 (12+log(O/H) $= 8.13 \pm 0.19$) and J1316+2614 (12+log(O/H) $= 8.45\pm0.12$) obtained in \cite{marques2020b} and \cite{marques2022}, respectively, using the R23 metallicity calibrator. The derived metallicities for the nine sources have mean value and scatter of 12+log(O/H) $=8.36\pm0.11$, that is, $Z/Z_{\odot}\simeq 0.5$ (taking solar value of 12+log(O/H) $=8.69$), and are listed in Table \ref{table_measurements}. Figure ~\ref{fig_met} shows the relationship between the strength of the He~{\sc ii} $\lambda 1640$ emission and 12+log(O/H). We find that sources with higher metallicities tend to have stronger \Heiiuv \ emission, suggesting a tentative correlation between (O/H) and $EW_{0}$ (\heii).

\subsection{No indications of AGN activity}

Finally, we investigate the presence and possible contribution of AGN activity since these sources are extremely bright in the UV, presenting apparent magnitudes similar to QSOs ($M_{\rm UV} \simeq -23.3$ to $\simeq -24.6$, see Table \ref{tab1}). In addition to the already mentioned P-Cygni wind lines, the GTC spectra of these sources also reveal photospheric absorption lines (see Appendix \ref{A1_GTC_spectra}), several of them detected with high significance. Examples of these are O~{\sc iv} $\lambda 1343$ or S~{\sc v} $\lambda 1501$. The identification of these inherently faint lines provides clear evidence that the UV luminosity is predominantly governed by starlight \citep[e.g.,][]{Gonzalez_Delgado_1998, shapley2003} as they originate from the photospheres of hot and massive stars. It is worth noting that even a minor contribution from an AGN to the UV continuum, which is featureless in these spectral regions, would result in the attenuation of these lines to a level imperceptible given the SNR of our spectra. As an additional test, we also look at the spectral profile of the Balmer lines in the EMIR rest-optical spectra. All sources show narrow profiles in Balmer lines (H$\beta$ or H$\alpha$) with intrinsic line widths between $\rm FWHM \lessapprox  400 \ km\, s^{-1}$ (unresolved) and $\simeq 470 \rm \ km\, s^{-1}$. This contrasts with the much broader line profiles of the non-resonant He~{\sc ii} $\lambda 1640$ which show $\rm FWHM \simeq 1150-2900 \ km\, s^{-1}$ (Table \ref{table_measurements}), clearly indicating stellar origin (WR and/or VMS). We thus conclude that the AGN contribution to the UV luminosity and strength of He~{\sc ii} in these sources is likely to be residual or null.

\section{Empirical analysis on the presence of VMS}
\label{ss_empirical_results}

\subsection{Evidence of VMS in UV-bright galaxies}

Several empirical arguments suggest a significant contribution of VMS in our sources, or at least in some of them, that we now describe. The most important one refers to the very high $EW_{0}$~(\heii) observed in our sources and the comparison with normal/typical star-forming galaxies where WR stars are expected. Indeed, WR stars are formed in basically all conditions and, thus, are expected to be present in all star-forming galaxies (of course, knowing that their contribution to the \heii \ line will be dependent on star-formation histories, age, metallicity, and other factors). In fact, broad \heii \ emission is recurrently observed in the rest-frame UV spectra of normal star-forming galaxies of enough S/N \citep[e.g.,][]{shapley2003, nolls2004, cabanac2008, mirka2010, jones2012, marques2020}, but its strength is significantly weaker than in our sources. 

To investigate the contribution of VMS and/or WR stars in our UV-bright galaxies, we first create a composite spectrum to show the resulting average spectral shape of \heii \ in our sources and compare it with the average rest-frame UV spectrum of normal LBGs from \cite{shapley2003} where WR stars are expected. First, the spectra of the 13 sources were de-redshifted using the systemic redshifts shown in Table \ref{tab1} and were resampled onto a common wavelength grid using linear interpolation. Next, we normalized the spectra using several spectral regions that are relatively free of emission or absorption features (i.e., excluding regions with ISM absorption, wind features, or nebular emission). Finally, we stacked all spectra by averaging the flux in each spectral bin.

\begin{figure*}
    \centering
    \includegraphics[width=\linewidth]{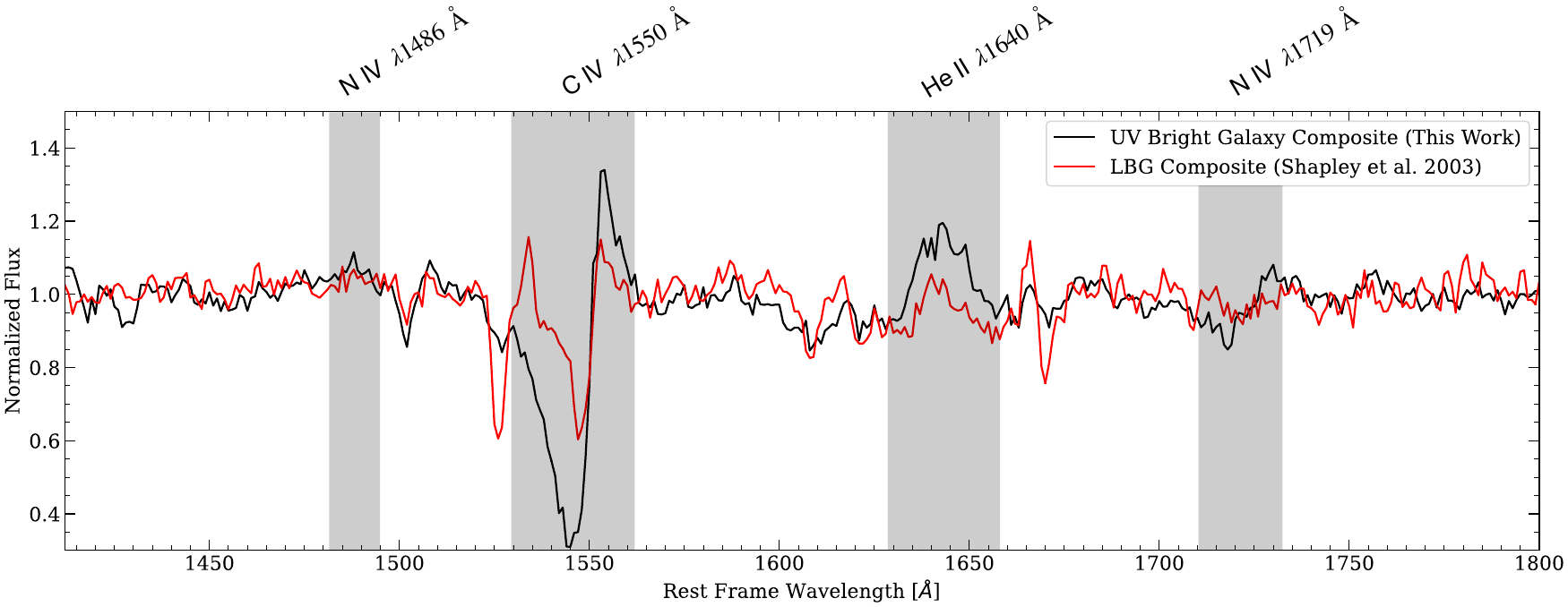}
    \caption{Comparison between our composite spectrum of UV-bright galaxies (black) and that of typical ($L_{\rm UV}^{*}$) Lyman break galaxies at $z \sim 3$ from \cite{shapley2003} (red). Measurements of the $EW_{0}$~(He~{\sc ii}) for both composites are shown in the caption. The region near the \Heiiuv \ emission profile is normalized with a different normalization factor to match the level of the continuum. The measured $EW_{0}$~(He~{\sc ii}) for the composite of UV bright galaxies in this work is 3.19 ± 0.07 \AA \ while for the Shapley composite is 1.37 \AA.}
    \label{Shapley_comparison_heii}
\end{figure*}

Figure \ref{Shapley_comparison_heii} shows the resulting stacked GTC spectrum of UV-bright galaxies and the comparison with that of typical ($L^{*}_{\rm UV}$) star-forming galaxies \citep{shapley2003}. As seen from Figure \ref{Shapley_comparison_heii}, the composite of UV-bright galaxies shows much stronger \heii \ emission than normal LBGs, by a factor of $\gtrsim 2$. Using the same integration windows as used in Section \ref{sec31} to derive the strength of \Heiiuv \ in our individual spectra, we measure $EW_{0}$~(\heii) $=3.19 \pm 0.07$ \AA \ for the stacked spectrum of UV-bright galaxies and $EW_{0}$~(\heii) $=1.37$ \AA \ for the \cite{shapley2003} composite. Since WR stars produce the bulk of the \heii \ emission observed in the \cite{shapley2003} composite spectrum, as quantitatively demonstrated by \cite{brinchmann2008pettini} and \cite{Eldridge2012}, we suggest that the excess of \heii \ observed in UV-bright galaxies is due to the contribution of VMS. Furthermore, the C~{\sc iv} P-Cygni appears much stronger in the composite of UV-bright galaxies than in normal LBGs. As shown later in Section \ref{ss_strength_nv}, VMS contribute significantly to the strength of C~{\sc iv} in the integrated spectrum.

The differences in the strengths of the He~{\sc ii} and C~{\sc iv} features seen in the composites of UV-bright and normal galaxies may also arise from differences in metallicities between these two galaxy populations, recognizing the metallicity dependence on the strength of the He~{\sc ii} and C~{\sc iv} wind lines. While precise measurements of the metallicity of the \cite{shapley2003} composite are not available, \cite{steidel2014} and \cite{Sanders2021} have obtained and studied the rest-frame optical spectra of similar galaxies (LBGs at $z \sim 2-3$). These works derive 12+log(O/H) between 8.2 to 8.6 which is broadly consistent with that obtained for most of our sources (Figure \ref{fig_met}). So, overall we expect the O/H abundance of the \cite{shapley2003} composite and our UV bright galaxies to be broadly similar, suggesting that the differences between the He~{\sc ii} and C~{\sc iv} profiles in our sources and normal LBGs are mainly due to the presence/absence of VMS.

In addition, a significant fraction of our sources show other spectral features that are present in the spectra of individual VMS (e.g., R136-a1, -a2, -a3, or R146 see: \citealt{crowther2016, brands2022, fabrice2022}). In particular, broad emission in N~{\sc iv} $\lambda 1486$ and a significant P-Cygni line profile in N~{\sc iv} $\lambda$1719 are clearly detected in J0006+2452, J0036+2725, J0110-0501, J0146-0220, J1157+0113, J1220+0842, and partially/barely detected in J0115+1837, J0121+0025, and J1335+4330 (see Appendix \ref{A1_GTC_spectra}). The detection of these profiles is also reflected in the composite spectrum of our UV-bright galaxies shown in Fig.~\ref{Shapley_comparison_heii}. These profiles are also seen in the integrated spectra of clusters or compact star-forming regions where VMS are suspected, including NGC 3125-A1 analyzed in \cite{Wofford2014, Wofford2023}, or J1129+2034, J1200+1343, and J1215+2038 analyzed by \cite{senchyna2017, senchyna2021}. Furthermore, these features are also predicted by the VMS models of \cite{fabrice2022}. However, we note that some WR-dominated clusters also show these spectral features, and thus they are not restricted to VMS  as pointed out by \cite{Martins2023}. As such, the \Nivuvnear \ and \Nivuvfar \ features could be also expected in the composite spectra of $L_{\rm UV}^{\star}$ LBGs, but these are faint or not detected as shown in Figure \ref{Shapley_comparison_heii} possibly due to a dilution effect from the contribution of the more numerous OB-type stars.

Finally, the last argument is related to the star-formation histories. The \heii \ line is boosted in bursty/instantaneous star-formation histories when compared to continuous star-formation histories, regardless of whether \heii \ originates from VMS or WR stars (see Section \ref{ss_models_results}). 
The rest-frame UV spectra of our sources show signs of smooth/continuous star-formation histories, as they show strong P-Cygni line profiles in both N~{\sc v} $\lambda 1240$ and Si~{\sc iv} $\lambda \lambda 1393,1402$ which are basically impossible to fit simultaneously with models of instantaneous bursts. In addition, SED analysis of the multiwavelength photometry and nebular emission performed by \cite{marques2020b, marques2021, marques2022} for a few of the sources studied here also suggest continuous star-formation histories, yet with very short ages ($\simeq 10$~Myr). 

\subsection{Spectral comparison of UV-bright sources with other VMS-dominated clusters}

\begin{figure*}
    \centering
    \includegraphics[width=\linewidth]{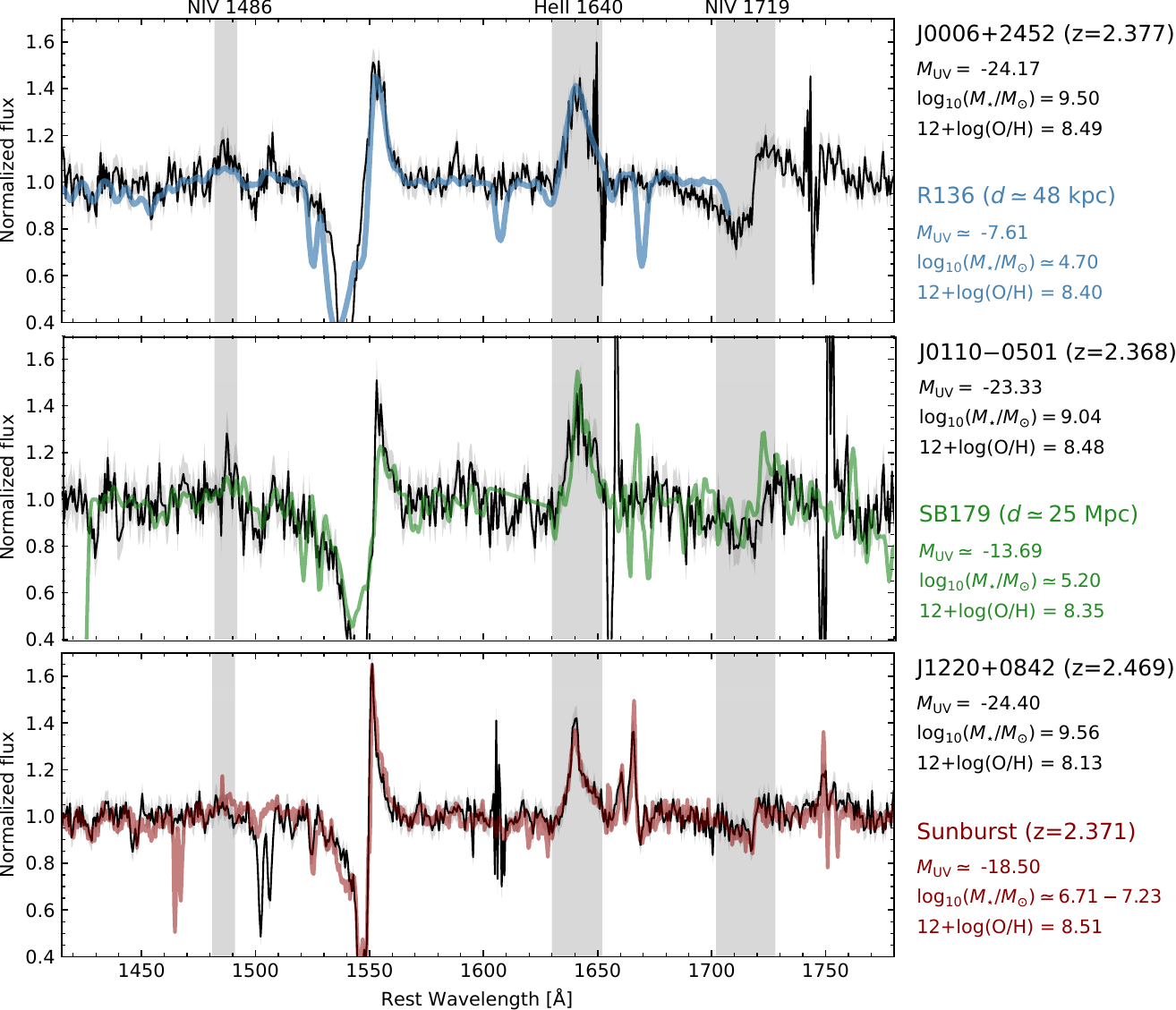}
    \caption{Comparison of the normalized GTC rest-UV spectrum of some of our sources with star clusters containing VMS (R136 in blue: \citealt{crowther2016}; SB 179 in green: \citealt{senchyna2017}; and the Sunburst cluster in red: \citealt{Mestric2023}). gray-shaded regions mark the location of the most important spectral features associated with VMS: broad emission in N~{\sc iv} $\lambda 1486$ and He~{\sc ii} $\lambda 1640$, and a P-Cygni line profile in N~{\sc iv} $\lambda 1719$. Some physical properties of these systems, including $M_{\rm UV}$, stellar mass, and metallicity, are also shown (see text for details). 
    }
    \label{fig_compasiron_VMS_clusters}
\end{figure*}

Here we present a comparative analysis of the rest-frame UV spectra of some of our UV-bright sources with those of known VMS-dominated clusters. Figure \ref{fig_compasiron_VMS_clusters} shows the spectra of J0006+2452 ($z=2.377$, $M_{\rm UV}=-24.17$, top), J0110-0501 ($z=2.368$, $M_{\rm UV}=-23.33$, middle), and J1220+0842 ($z=2.469$, $M_{\rm UV}=-24.40$, bottom). Overlaid are the spectra of three young star-clusters where the presence of VMS is confirmed or suspected. Specifically, we include the spectrum of the R136 cluster within 30Dor/LMC (\citealt{crowther2016}, blue), SB 179 (\citealt{senchyna2017}, green), and the highly magnified Sunburst cluster at  $z\simeq 2.371$ (\citealt{Mestric2023}, red). This comparative analysis serves primarily for illustrative purposes, aimed at highlighting the resemblances between some of our UV-bright galaxies and established examples of VMS-dominated clusters.

Overall, the spectra of our sources resemble those of VMS-dominated clusters. In particular, the spectral features typically associated with WN and WNh stars (including VMS), such as the intense and broad He~{\sc ii}~$\lambda 1640$ emission and the N~{\sc iv} profiles at $\lambda 1486$ and $\lambda 1719$ (highlighted in gray regions), are present in both samples presenting also similarities in their spectral shapes. A closer examination also reveals differences. For example, the N~{\sc iv} $\lambda 1486$ appears stronger in J0006+2452 and J0110-0501 than in R136 and SB 179, although effects of different resolution and SNR between spectra prevent a fair comparison. The blueshifted absorption component of the N~{\sc iv}~$\lambda 1719$ P-Cygni line profile is also more pronounced in J0006+2452 and J0110-0501, showing substantially larger terminal velocities than what is seen in the spectra of R136 or SB 179. On the other hand, the He~{\sc ii} profiles are almost identical between our sources and the comparison sources. Of course, variations in the spectral shapes and intensities of He~{\sc ii} or N~{\sc iv} profiles or others are expected and can be attributed to factors such as age, metallicity, IMF, or other intrinsic properties of massive and very massive stars (e.g., mass-loss, terminal wind velocities, rotation, etc.). As already highlighted in Fig.~\ref{HeII_profiles}, the spectra of our 13 sources exhibit diverse He~{\sc ii} profiles, ranging from nearly Gaussian/symmetric profiles (e.g., J0006+2452 or J0115+1837, although with different line widths) to asymmetric profiles (J1220+0842 or J1335+4330), or more complex ones with multiple emission/absorption peaks (e.g., J0031+3545 or J0121+0025). Similarly, the spectra of nearby star-forming regions analyzed by \cite{senchyna2021} or \cite{Martins2023} also reveal differences in the shape in the He~{\sc} line (and N~{\sc iv} as well) from source to source (VMS- or WR-dominated sources).

\subsection{Empirical classification of VMS- or WR-dominated sources}

\cite{Martins2023} provide observational criteria to distinguish sources dominated by VMS or WR stars. In their scheme, VMS-dominated sources show intense and broad \Heiiuv \ emission, with $EW_{0} \geq 3.0$ \AA{ }at least. This limit is motivated by the empirical predictions of \cite{schaerer1998} of the intensity of the \Heiiuv \ line from various types of WR stars in young stellar populations, which reach a maximum of $EW_{0} \simeq 3.0$ \AA \ for an instantaneous burst at $Z/Z_{\odot} \simeq 0.5$. 
Furthermore, the spectra of VMS-dominated sources also present broad optical He~{\sc ii}~$\lambda 4686$, but on the other hand, the emission in N~{\sc iii}+C~{\sc iii} 4640\AA-4650\AA \ and C~{\sc iv} 5801\AA-5812\AA, should be weak or absent \citep{Martins2023}. Given that our rest-optical spectra are too shallow to detect any of these optical lines or they are not available at all, we restrict our classification to the strength of the 1640 \AA \ line. 

Following the observed strengths in the \Heiiuv \ line in our sources (see Table \ref{table_measurements}), 8 of them show $EW_{0}$~(\heii) $>3.0$ \AA \ and thus are likely to be dominated by VMS. An additional source, J1220+0842, represents a special case. Its $EW_{0}$~(\heii) $=2.52 \pm 0.33$ \AA \ falls below our selection threshold. However, the source shows strong and broad N~{\sc iv} $\lambda$1486 and a P-Cygni line in N~{\sc iv} $\lambda$1719, both of which are also VMS signatures. This galaxy also has the lowest metallicity in our sample (12+log(O/H)$=8.13 \pm 0.19$, Figure \ref{fig_met}), which might be expected to weaken the \heii \ line (due to weaker stellar winds). Given its spectral similarity to the Sunburst cluster and others \citep[][see Figure \ref{fig_compasiron_VMS_clusters}]{Mestric2023}, we add this source to our sample of candidate VMS hosts.

In summary, we conclude that among the 13 UV-bright galaxies analyzed in this work, nine of them are likely VMS-dominated sources. We conservatively consider them as candidates for the reasons described above, and in particular due to the lack of deep observations in the rest-optical blue and red bumps, which according to \cite{Martins2023} are necessary to unambiguously distinguish VMS- and WR-dominated sources. The three strongest He~{\sc ii} emitters in our sample, J0006+2452 ($EW_{0} = 4.54$\AA), J0110$-$0501 ($EW_{0} = 4.57$\AA), and J1316+2614 ($EW_{0} = 4.48$\AA) stand out to be the best VMS candidates in our sample. The observed $EW_{0}$~(He~{\sc ii}) in these sources are similar to those in the spectra of local star-clusters or star-forming regions dominated by VMS (e.g., R136, NGC 3125-A1, J1129+2034/SB179, \citealt{crowther2016, senchyna2021, Wofford2023}), even noting that these local clusters may have their He~{\sc ii} emission enhanced by bursty/single-age star-formation histories.

\section{Population synthesis models}
\label{ss_models_results}

Population synthesis models can be used to infer stellar populations from the comparison of synthetic and observed UV spectra of star-forming galaxies. This approach was applied in a few local star clusters like NGC 3125-A1, NGC 5253-5, or II~Zw~40, where intense \Heiiuv \ emission is detected ($EW_{0} \gtrsim 4.0$ \AA; \citealt{Wofford2023, Wofford2014, Leitherer2018, Smith2016}). The strength of \Heiiuv \ observed in these clusters is well above that any population synthesis models without VMS can predict, leading these authors to suggest the presence of VMS. For these reasons, it is important to discuss first the validity and limitations of different synthesis models with/without VMS in predicting the \Heiiuv \ emission.
In this Section, we investigate the ability of various population synthesis models to account for the strong UV lines of our sources. We consider models available in the literature and apply new models of VMS from \citet{fabrice2022}.

\subsection{BPASS models}
\label{ss_bpass}

We first retrieve the spectral energy distributions (SEDs) produced by BPASS v2.2.1 models \citep{bpass2017, bpass2018, bpass2022}. We use BPASS models with binary stellar populations and upper mass limits of 100 $M_\odot$ and 300 $M_\odot$ with a Salpeter IMF slope of $-2.35$. Furthermore, we select a metallicity of $Z=0.006$ since it is the closest to the average metallicity of these sources (12+log(O/H) $=8.36\pm0.11$, Section \ref{s_sources}) and also the closest to the metallicity of the Large Magellanic Cloud (LMC) for which the new VMS models of \cite{fabrice2022} were developed. This is also consistent with the previous works of \citet{marques2020b, marques2021, marques2022} on a subset of our sources. 

BPASS models have instantaneous star formation histories (ISFH) normalized to a burst mass of $10^6 \ M_\odot$. The detection of strong P-Cygni line profiles in both N~{\sc v} $\lambda 1240$ and Si~{\sc iv} $\lambda 1400$ in the spectra of these sources, which originate respectively from O-type main-sequence and supergiant phases, is difficult to be explained by single ISFH model (e.g., see \citealt{chisholm2019}) and suggests a more smooth/continuous star-formation history. Hence we convert the SEDs of ISFH models to constant star formation history (CSFH) models, which we expect in our sources. We discuss how the models perform at reproducing the strength of the main UV lines in Sect.~\ref{ss_strength_nv} to \ref{ss_strength_nv_heii}.

\subsection{Models including VMS self-consistently}
\label{ss_vms}

\subsubsection{Stellar models}
 
Spectra of four single-star VMS are presented in \cite{fabrice2022} at masses of 150, 200, 300, and 400 $M_\odot$, and were computed using nonrotating evolutionary models using the code {\sc STAREVOL} \citep{siess2000, Lagarde2012, Amard2019}. VMS models use the mass loss prescription of \cite{grafener2021} which are developed for a metallicity of 0.4 $Z_\odot$, considering $Z_\odot$ = 0.0134. The model traces the evolution of each VMS from the zero-age main sequence to the end of the H-burning phase which lasts somewhere from 2 to 2.5 Myr depending on the birth mass of the VMS.

In these VMS models, the proximity of the star to the Eddington limit makes the scaling of mass loss rates to the Eddington factor steeper, which differs from the prescription used in \cite{vink2001}. This leads to a higher mass loss rate with optically thicker winds. It results in higher helium surface abundance which, combined with higher wind density, produces stronger \Heiiuv \ emission. The \heii \ appears as a P-Cygni line profile that strengthens with age (see, \citealt{fabrice2022}). Apart from \Heiiuv, the \Nvuv \ and \Civuv \ profiles also appear as P-Cygni line profile. 

Other important features that show up in the theoretical UV spectra of VMS are the \Nivuvfar \ and \Nivuvnear \ profiles. The \Nivuvfar \ appears weak at the ZAMS but evolves into a stronger P-Cygni line profile with age. The \Nivuvnear \ profile appears as an emission only after 1.5 million years of evolution of the VMS and becomes stronger with age. The \Nivuvnear \ profile is not seen in the less massive O-type supergiants even at solar metalicity \citep{walborn1985, Bouret2012} but it appears on the UV spectra of WN and WNh stars \citep{Hamann2006, Hainich2014}.  The stellar emission in \Nivuvnear \ appears broad (FWHM $>1000$ km~s$^{-1}$) in VMS models \citep{fabrice2022} and in VMS-dominated clusters (e.g., R136; \citealt{crowther2016}) and can be distinguishable from the nebular emission observed in some narrow-lined UV-selected AGNs \citep[e.g.,][]{hainline2011b} and star-forming galaxies. This emission feature only rarely occurs in broad-lined quasars in the SDSS \citep[see,][]{Bentz2004b, Jiang2008}, and is also detected in a few radio galaxies \citep[see,][]{Vernet2001, humphrey2008}. The \Nivuvfar \ P-Cygni line profile has been observed in some hot Of-type stars \citep[e.g.,][]{conti1996}.

BPASS v2.2.1 models extending to 300 $M_\odot$ do incorporate evolutionary tracks for both single and binary very massive stars. However, these adopt the same wind mass loss rates in the VMS regime as for massive O-Stars \citep{vink2001}. The BPASS spectral synthesis also assigns these stars template spectra derived from the same stellar atmosphere template grid as other hot, massive stars (derived from the WMBasic atmosphere code), rather than producing custom spectra in this regime to reflect the different wind regime. Recent works on VMS have suggested that both the mass loss prescription and the resulting atmospheric radiative transfer of VMS might differ substantially from hot stars below 100 $M_\odot$ \citep[see e.g.,][]{grafener2021, Bjorklund2023}.

\subsubsection{Producing synthetic population spectra with VMS}
\label{ss_vmsbpass}

To describe the integrated UV spectra of star-forming populations we combine the SEDs of VMS models from \cite{fabrice2022} that are available for VMS of 150, 200, 300, and 400 $M_\odot$ to normal stellar populations from the BPASS v2.2.1 models that include stars up to 100 $M_\odot$; the method is similar to that adopted in \cite{fabrice2022}. We have defined mass bins in the range [100, 175], [175, 225], [225, 275], [275, 325], and [325, 475] and calculated the number of VMS required in these mass bins by extrapolating the IMFs to the upper bin mass. In this approach, we have used a corrected equation for a continuous IMF, which produces significantly different numbers than that has been used in \cite{fabrice2022}. The details of this process is described in Appendix \ref{imf_extrapolation}. In short, we take the BPASS models up to 100 $M_\odot$ and assume the same IMF (Salpeter) for stars with higher masses. Then, the VMS spectra are resampled to a common wavelength grid using the {\sc Spectres}\footnote{\url{https://spectres.readthedocs.io/en/latest/}} \citep{spectres2017} \ python library. We consider five different maximum masses of VMS, between 175 and 475 $M_\odot$. This is done for all burst models with ages up to 2.5 Myr, after which no VMS is left in the population. The ISFH models including VMS are then also used to compute models for constant SFR. We note that given the assumptions of \cite{fabrice2022}, the post-main sequence evolution of VMS is neglected here, and we also neglect VMS in binary systems.

\subsubsection{Validation of synthesis models with/without VMS}

The VMS models used in this work were introduced by \cite{fabrice2022}. When coupled with synthetic population spectra from BPASS, \cite{fabrice2022}  demonstrated that their models successfully explain the strength of \Heiiuv \ observed in the R136 cluster ($EW_{0} \simeq 4.5$\AA), which is produced primarily from the few VMS located in the core of R136 \citep{crowther2016}. Other synthetic population spectra including VMS like the new Bruzual \& Charlot models \citep{plat2019}, which incorporate VMS to $300 M_{\odot}$, have been also used in studies to investigate the presence of VMS in local star-forming regions \citep[e.g.,][]{senchyna2021, Wofford2023, Smith2023}. Even noting that  Bruzual \& Charlot and \cite{fabrice2022} models have different VMS prescriptions (see references therein for details), they successfully reproduce several UV spectral features observed in VMS, and in particular, by boosting the \Heiiuv \ strength in the integrated synthetic population spectra. We thus conclude that the VMS models used in this work are valid to search for VMS in our sources. 

A different situation may apply to standard synthetic spectra without VMS, where \heii \ is expected to be produced primarily by classical WR stars. Figure \ref{fig_ew_with_age} shows that the BPASS models with the upper mass cutoff of $100 M_{\odot}$ (i.e., without VMS) struggle to exceed 1 \AA \ and 2 \AA \ in \Heiiuv\ for continuous and instantaneous star formation histories, respectively, at the adopted metallicity $Z=0.006$. These are slightly lower than those found in \cite{brinchmann2008pettini} using Starburst99 continuous star formation models at similar metallicity, but including also the contribution of Of stars (reaching a maximum of $EW_{0} \simeq 1.4$ \AA), or the burst models of \cite{schaerer1998} that rely on empirical line luminosities of different types of WR stars and reach a maximum of $EW_{0} \simeq 3.0$ \AA \ (at $Z/Z_{\odot} \simeq 0.5$). Furthermore, the UV spectroscopic analysis by \cite{Martins2023} of seven WR-dominated local star-forming regions at $Z\simeq 0.5 Z_{\odot}$ show somehow stronger \Heiiuv \ emission than the ones measuring here using BPASS models. They measure strengths in the 1640 \AA \ line of $\simeq 1.64-2.35$ \AA \ for six of them, and an extreme value of $EW_{0}$~(\heii) $\simeq 3.89$ for Tol 89-A, a known local WR-dominated cluster \citep{sidoli2006}. Adopting a metallicity of 0.008 instead of 0.006 for the M$<100~M_{\odot}$ BPASS population only marginally increases the \heii\ emission. From these results and the analyses of the optical blue and red bumps, \cite{Martins2023} suggested that the treatment of WR stars should be improved in current population synthesis models (see further details in  \citealt{martins2023b}).

We now examine the predicted UV spectra from synthesis models with and without VMS and compare them to our observations. To do so we use the models for constant star-formation, except if otherwise stated.

\subsection{Strength of N~{\sc v}~$\lambda 1240$, and C~{\sc iv}~$\lambda 1550$ \ P-Cygni line profiles}
\label{ss_strength_nv}

\begin{figure*}[h]
    \centering
    \includegraphics[width=0.94\linewidth]{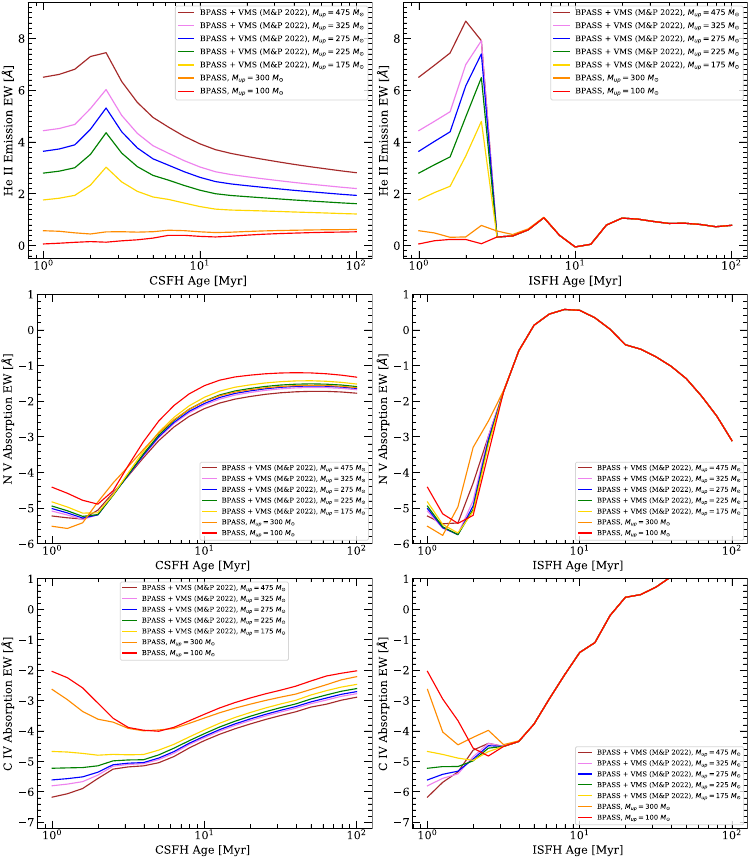}
    \caption{Variation of the equivalent width ($EW_{0}$) of the absorption component of the \Nvuv \ (middle), \Civuv \ (bottom) P-Cygni line profiles and the stellar \heii \ $1640$ \AA \ emission (top) as a function of age assuming a continuous star formation history (left) and instantaneous burst models (right). Different tracks represent models with different IMF upper mass limits ($M_\mathrm{up}$). Original BPASS models with $M_{\rm up} =100 M_{\odot}$ and $300 M_{\odot}$ (using the wind prescription from \citealt{vink2001}) are shown in red and orange, respectively, while BPASS models with the new VMS template spectra of \cite{fabrice2022} are shown in yellow, green, blue, pink, and brown for $M_\mathrm{up} =175 M_{\odot}$, $225 M_{\odot}$, $275 M_{\odot}$, $325 M_{\odot}$, and $475 M_{\odot}$ respectively.}
    \label{fig_ew_with_age}
\end{figure*}

The rest-frame UV spectra of our sources (Appendix \ref{A1_GTC_spectra}) show indications of the presence of young and massive stars. In particular, strong P-Cygni line profiles are observed in \Nvuv \, and \Civuv \ which are sensitive to age and metallicity in the UV \citep[see e.g., ][]{chisholm2019}. Massive stars and VMS are very hot stars with surface temperatures ranging from 40,000 K to 60,000 K. As such, the dominant ionization state of stellar wind changes from $N^{3+}$ to $N^{4+}$, at a given metallicity \citep[see,][]{kudritzki1998, lamers1999, leitherer2010}. This results in a strong P-Cygni line profile with a blueshifted absorption and a redshifted emission. As stars evolve to relatively older ages the dominant ionization state changes to $N^{3+}$ resulting in a decrease in the strength of \Nvuv \ P-Cygni line profile, that is, decreasing the strength of the corresponding absorption and emission. This makes the strength of the \Nvuv \ P-Cygni line profile a potential tracer of the age of the stellar population in young star-forming galaxies. 

We perform a quantitative study on the strength of \Nvuv \ P-Cygni line profile by measuring the strength of the absorption component of the \Nvuv \ P-Cygni profile, which is the most affected by the change in age both in our sources and models \citep[][]{chisholm2019}. We measure the $EW_{0}$ of the N~{\sc v} absorption in the spectral window from 1230 \AA \ to 1240 \AA, avoiding as much as possible the contamination from the Ly$\alpha$ emission, the left-most part of this integration is chosen to be around $\sim$ 3500 \ km s$^{-1}$ away from the Ly$\alpha$ rest-wavelength. However, due to nebular Ly$\alpha$ emission from the galaxy, which is not accounted for by our models and which can extend redward by several hundred km s$^{-1}$, we refrain from using the \Nvuv \ absorption as an absolute age indicator. For the estimation of the continuum level, we select the spectral region from 1269 \AA \ to 1272.5 \AA \ to avoid the contribution of the low-ionization Si~{\sc ii} $\lambda1260$ absorption line. 

The middle panels of Figure~\ref{fig_ew_with_age} show how the \Nvuv \ absorption strength varies as a function of age for models assuming a CSFH (left) and ISFH (right) with different upper mass limits of the IMF ($M_{\rm up} = 100-475 M_{\odot}$). At a very young ages ($\sim$ 1 Myr), \Nvuv \ shows very strong absorption, between $EW_{0} \simeq -4.4$ \AA \ for a non-VMS IMF (i.e., $M_{\rm up}=100 M_{\odot}$, red curve in Fig. \ref{fig_ew_with_age}) and $EW_{0} \simeq -5.6$ \AA \ when VMS are included. At later ages ($\gtrsim 2$Myr), the strength of the absorption component of N~{\sc v} decreases, reaching a plateau of $EW_{0} \simeq -1.5$ \AA \ at $\gtrapprox 20$ Myr for CSFH models. This figure also shows that the inclusion of VMS provides only a marginal contribution to the \Nvuv \ absorption strength at any age, although this effect is more significant at $\lesssim 2$ Myr. It is also worth noting that the original BPASS models with $M_{\rm up} = 300 M_{\odot}$ (orange curve in Fig.  \ref{fig_ew_with_age}) and the ones using VMS from \cite{fabrice2022} predict similar strengths of \nv, even noting that they assume different wind prescriptions: \cite{vink2001} for the former, and \cite{grafener2021} for the latter. Applying the same methodology to our sources, we measure $EW_{0}$ of the \nv \ absorption between $\simeq -5.5$ \AA \ to $\simeq -1.1$ \AA \ (Table \ref{table_measurements}), indicating different ages according to Fig.~\ref{fig_ew_with_age}. In the absence of any contamination by the Ly$\alpha$ emission, the sources J0006+2452, J0036+2725, and J0110-0501 have the strongest absorption in \nv \ with $EW_{0} \lesssim -3.5$ \AA \ and, therefore, stand out to be the youngest systems in our sample. The rest of the sources appear slightly older, in particular, those with $EW_{0} \gtrsim -3.0 $ \AA.

We perform a parallel quantitative investigation on the \Civuv \ P-Cygni profile, measuring its absorption strength. Given the presence of the interstellar medium (ISM) component in \Civ, we measure the $EW_{0}$ of the stellar \Civuv \ absorption within the spectral range spanning from 1530 \AA \ to 1543 \AA, that is, at least $\sim$ 1000 km s$^{-1}$ away from the rest-velocity of C~{\sc iv} $\lambda 1548$. Even with the selection of these spectral windows, we cannot rule out the ISM contamination in the stellar \Civuv \ absorption, in particular in the presence of strong outflows. To establish the continuum level, we designate the spectral region from 1509 \AA \ to 1519 \AA \ on the left and a region from 1570 \AA \ to 1580 \AA \ on the right side of the \Civuv \ profile. For J1316+2614, the continuum window on the right is exclusively selected from 1578 \AA \ to 1580 \AA \ due to atmospheric line contamination in the optical GTC spectra.

The bottom panels of Figure~\ref{fig_ew_with_age} show how the \Civuv \ absorption strength varies as a function of age for models assuming a CSFH (left) and ISFH (right) with different upper mass limits of the IMF ($M_{\rm up} = 100-475 M_{\odot}$). Unlike \Nvuv \ absorption profile, the \Civuv \ absorption is susceptible to different mass loss prescriptions. The \cite{grafener2021} mass loss prescriptions used in the \cite{fabrice2022} models significantly boost \Civuv \ absorption; for the CSFH model with IMF $M_{\rm up} = 300 M_{\odot}$ at $\sim$ 1 Myr, its strength is boosted by 2.4 \AA \ in comparison to the BPASS model at the same age and IMF. At this age, the C~{\sc iv} shows very strong absorption between $EW_{0} \simeq -4.7$ \AA \ to $-6.2$ \AA \ for models that include VMS with \cite{grafener2021} mass loss prescription. The strength only decreases slightly till 2.5 Myr. While the BPASS model with IMF $M_{\rm up}=300 M_{\odot}$ that includes \cite{vink2001} mass loss prescription, the strength starts with $EW_{0} \simeq -2.8$ \AA \ and reaches its strongest at $EW_{0} \simeq -4$ \AA \ at 2.5 Myr before reducing again.

\begin{figure*}
    \centering
    \includegraphics[width=0.98\linewidth]{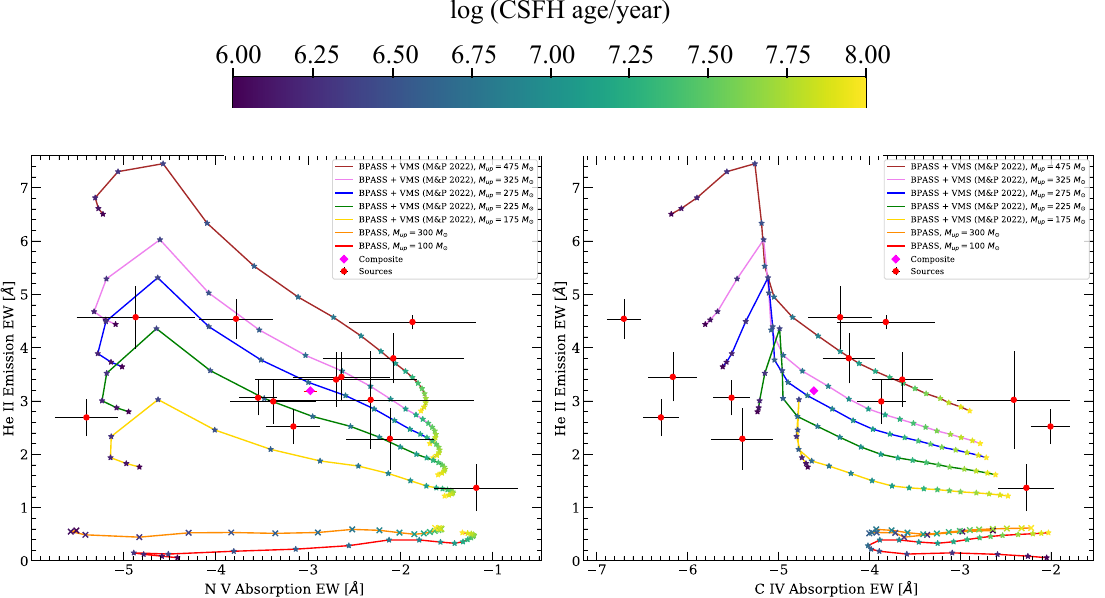}
    \caption{Variation of the $EW_{0}$ of \heii \ $1640$ \AA \ emission (y-axis) with the absorption component of the \Nvuv \ P-Cygni line profile (x-axis) on the left and the same with the absorption component of \Civuv \ P-Cygni line profile (x-axis) on the right as a function of age assuming a continuous star formation history. Measurements of our UV-bright galaxies are marked with solid circles. Different tracks represent models with different IMF upper mass limits ($M_{\rm up}$). Original BPASS models with $M_{\rm up} =100 M_{\odot}$ and $300 M_{\odot}$ (using the wind prescription from \citealt{vink2001}) are shown in red and orange, respectively, while BPASS models with the new VMS template spectra of \cite{fabrice2022} are shown in yellow, green, blue, pink, and brown for $M_{\rm up} =175 M_{\odot}$, $225 M_{\odot}$, $275 M_{\odot}$, $325 M_{\odot}$, and $475 M_{\odot}$ respectively.}
    \label{HeII vs NV}
\end{figure*}

We measure the C~{\sc iv} absorption strength for sources between $EW_{0} \simeq -6.7$ \AA \ to $-2.0$ \AA \ (Table \ref{table_measurements}). We can not rule out the contamination from the ISM in the C~{\sc iv} for our sources, and the contamination results in the C~{\sc iv} absorption strength being overestimated in our sources. But, the degeneracy related to metallicity has also been affecting the strength of C~{\sc iv} absorption in our sources, as mentioned earlier, it is a metallicity-sensitive feature in the UV spectra. The two sources J1220+0842 and J0850+1549 on the right-most end of Figure \ref{HeII vs NV} with lowest C~{\sc iv} absorption strength are the lowest metallicity sources in our sample with 12+log(O/H) $\simeq$ 8.13 and 8.27 respectively. While the sources with stronger C~{\sc iv} absorption strength have relatively higher metallicities.

\

\subsection{Strength of He~{\sc ii}~$\lambda 1640$ emission: VMS indicator}
\label{ss_strength_heii}

VMS have been already identified and characterized in detail throughout UV and optical spectroscopy, either individually \citep[e.g.,][]{Massey1998, Bestenlehner2014, crowther2016} or in integrated spectra of unresolved star-forming regions \citep[e.g.,][]{Wofford2014, Smith2016, Mestric2023, Wofford2023, Smith2023}. 
From these empirical results and models, broad and intense \Heiiuv \ emission appears to be ubiquitous in VMS, making it the best indicator of the presence of VMS in the rest-frame UV (see \citealt{fabrice2022}). Since classical WR stars can also produce broad \heii \ emission, the strength of \heii \ depends thus on the relative contribution of VMS and WR stars in integrated light spectra.

The \Heiiuv \ emission equivalent width has been measured by using both Gaussian fitting and flux integration methods, as shown in Fig. \ref{HeII_profiles}. However, the \heii \ profiles exhibit a non-Gaussian nature in the majority of our sources, while manifesting asymmetrical P-Cygni features in the models. To compute the \Heiiuv \ emission equivalent width for both the sources and the models, we have employed the flux integration method, as described in Section \ref{sec31}. The top panels of Fig.~\ref{fig_ew_with_age} show the variation of the equivalent width of the stellar \heii \ emission as a function of age for CSFH (left) and ISFH (right) models assuming different IMF upper mass limits. As shown in this figure, BPASS models with $M_{\rm up} = 100 M_{\odot}$ predict the weakest \heii \ intensity within our different models, with a maximum of $EW_{0}$(\heii) $\simeq 0.6$ \AA \ at $\gtrsim 6-8$~Myr for CSFH, that is, when the contribution of WR stars reaches its maximum. The original BPASS model with an upper mass limit of 300 $M_{\odot}$ (in orange), which has the wind prescription from \cite{vink2001}, can only predict a maximum of $EW_{0}$(\heii) $\simeq 0.8$ \AA. For comparison, earlier models of \cite{brinchmann2008pettini} predict a maximum  $EW_{0}$(\heii) $\simeq 1.4$ \AA\ for CSFH at similar metallicity. On the other hand, BPASS coupled with VMS models of \cite{fabrice2022} show much stronger \heii \ emission, with $EW_{0}$(\heii) ranging from  $\simeq 1.8-7.6$ \AA, depending on the age and $M_{\rm up}$. For these models, the \heii \ strength peaks near 2.5 Myr, approximately the lifetime of the VMS. From Fig.~\ref{fig_ew_with_age}, it is also evident that the impact of VMS in integrated spectra is much stronger in the \heii \ emission (boosting it by a maximum factor of $\approx 7.5$) than in the strength of \nv \ P-Cygni line profile (left panel of Fig.~\ref{fig_ew_with_age}).

\subsection{He~{\sc ii}~$\lambda 1640$ against N~{\sc v}~$\lambda 1240$ and C~{\sc iv}~$\lambda 1550$: Evidence of VMS}
\label{ss_strength_nv_heii}

Putting together the results obtained in Sections \ref{ss_strength_nv} and \ref{ss_strength_heii}, we show in Fig.~\ref{HeII vs NV} the relationship between the strength of the absorption component of the \nv \ and C~{\sc iv} \ P-Cygni line profiles and the stellar \heii \ emission. Figure \ref{HeII vs NV} also shows the measurements of $EW_{0}$(N~{\sc v}) and $EW_{0}$(He~{\sc ii}) of our sources (listed in Table \ref{table_measurements}). Figure \ref{HeII vs NV} provides strong evidence for the presence of VMS in most of our sources. The sources that lie on the left-hand side of this figure, J0006+2452, J0036+2725, and J0110-0501, show $EW_{0}$(N~{\sc v}) $\lesssim$ -4.5 \AA \ and are the youngest systems in our sample. These sources show $EW_{0}$(He~{\sc ii}) $\simeq 2.7-4.6$ \AA, requiring a significant number and contribution of VMS, or IMF upper mass limits between $M_{\rm up}=225$ M$_{\odot}$ and $M_{\rm up}=275$ M$_{\odot}$. For the remaining sources showing $EW_{0}$(N~{\sc v}) $\simeq$ -4.5 \AA \ to $\simeq$ -1.0 \AA \ and $EW_{0}$(He~{\sc ii}) $\simeq 2.0 - 4.5$ \AA, stellar population with VMS are also clearly preferred and possibly required, although the IMF upper mass limit is not well constrained given the large uncertainties in our measurements. An exception may be J1316+2614, showing $EW_{0}$(N~{\sc v}) $\simeq$ -1.9 \AA \ and $EW_{0}$(He~{\sc ii}) $\simeq 4.5$ \AA, which requires an IMF upper mass limit of $M_{\rm up}\gtrsim 475$ M$_{\odot}$.

\vspace{1mm}

Overall Fig.~\ref{HeII vs NV} shows that different BPASS + VMS models of \cite{fabrice2022} can predict relatively well both the observed strength of \Heiiuv \ emission and \Nvuv \ absorption of these sources. On the other hand, models without VMS are not able to reproduce the strength of \Heiiuv \ emission and \Civuv \ absorption. In line with the models shown in Fig.~\ref{HeII vs NV}, most of the sources in our sample require a significant contribution of VMS in integrated stellar populations to explain the observed strengths of wind lines, in particular \heii.

\subsection{Spectral comparison}
\label{ss_spectra_comp}

In this section, we have compared the strengths of the observed \heii \ stellar emission in our sources to those from different synthetic models. We now perform a qualitative comparison of the spectral profiles of the \heii \ line between our sources and models.

\begin{figure*}
    \centering
    \includegraphics[width=\linewidth]{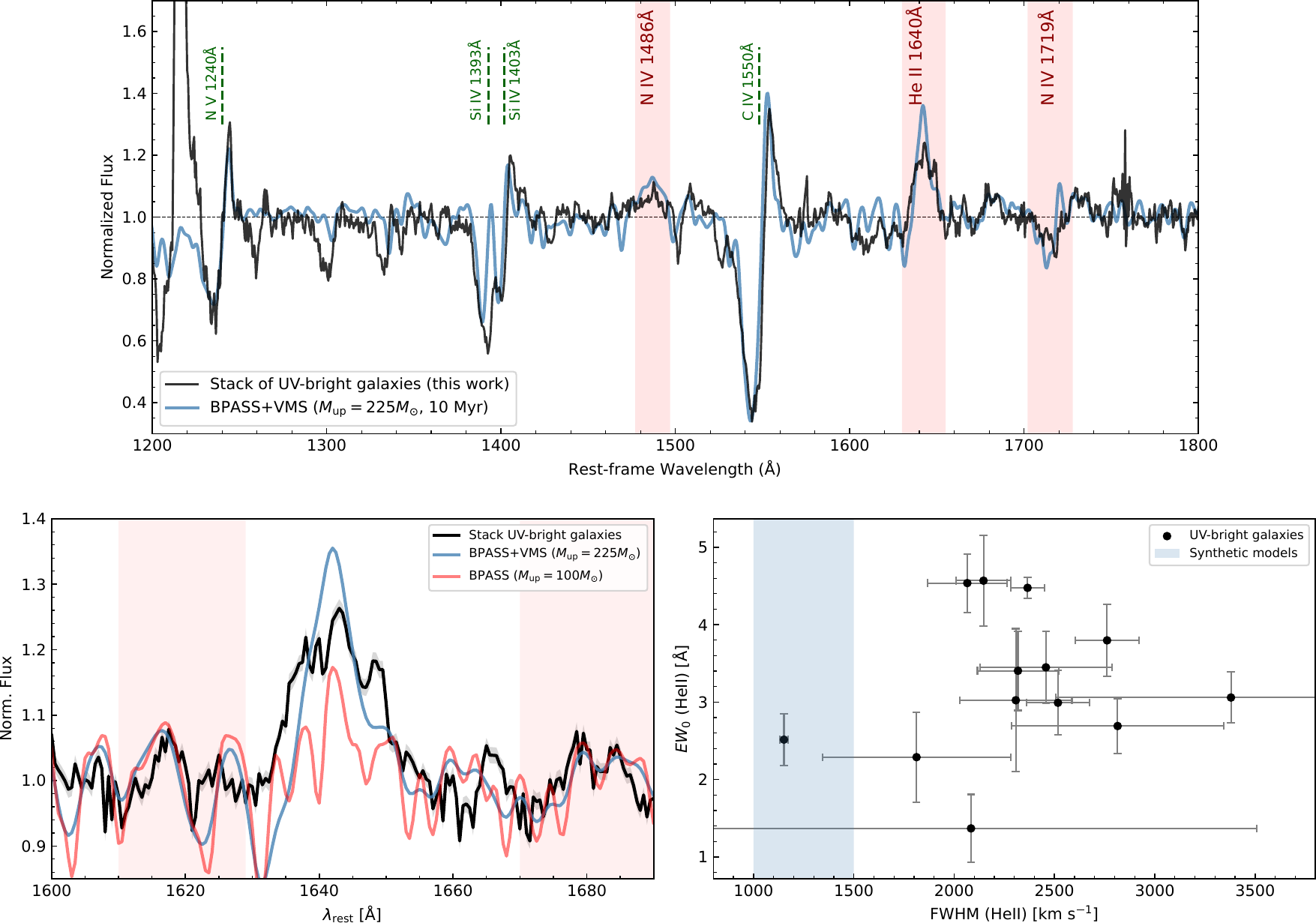}
    \caption{Top panel shows the normalized composite spectrum using the 13 UV-bright sources used in this work (black). The synthetic spectrum of a BPASS+VMS model is also shown in blue and assumes an upper mass cutoff of the IMF of $M_{\rm up} = 225 M_{\odot}$ with a continuous star-formation over 10 Myr (using the VMS models of \citealt{fabrice2022}). Spectral features associated with only VMS are marked in red while the spectral features produced by VMS and normal massive stars are marked in green. The bottom left panel shows a zoom-in at the \heii \ line. BPASS models with and without VMS are also shown (blue and red, respectively). The bottom right panel illustrates the differences between the \heii \ line widths (FWHM) measured in our sources (black) and model spectra (blue region). }
    \label{fig_matched_spec}    
\end{figure*}

The top panel of Fig.~\ref{fig_matched_spec} shows the resulting stack spectrum of our 13 UV-bright galaxies. The most prominent VMS features are highlighted in red (\Heiiuv \ and the N~{\sc iv} profiles at 1486\AA{ }and 1719\AA), as well as other stellar wind features produced by normal massive stars and VMS (N~{\sc v} $\lambda 1240$, Si~{\sc iv} $\lambda\lambda 1393,1402$, and C~{\sc iv} $\lambda 1550$). Using the same methodology described in Sections \ref{ss_strength_nv} and \ref{ss_strength_heii}, we measure the $EW_{0}$ of the absorption component of the N~{\sc v} P-Cygni line profile and the stellar emission of He~{\sc ii} of the stacked spectrum, finding $EW_{0}$~(N~{\sc v})~$=$-3.0 $\pm$ 0.07 \AA\ and $EW_{0}$~(He~{\sc ii})~$=3.19 \pm 0.07$ \AA, respectively. We also show in Fig.~\ref{fig_matched_spec} the normalized spectrum of a BPASS+VMS model \citep{fabrice2022} assuming an IMF with an upper mass cutoff of $M_{\rm up} = 225 M_{\odot}$ with a continuous star-formation over 10 Myr (blue). The model spectrum was also convolved to match the spectral resolution of the GTC spectra ($R\sim 700$). Overall, the synthetic spectrum can reproduce several important features associated with massive stars (e.g., N~{\sc v}, Si~{\sc iv}, or C~{\sc iv}). The spectral features related to VMS are also relatively well-reproduced by the model spectra, but differences between their spectral shapes are also evident. Given the fact that the spectrum of the composite of UV-bright galaxies is well reproduced by a constant star formation history model with $\approx 10$ Myr, it is more representative of a VMS+WR spectrum, where the contribution from the VMS in the rest-frame UV spectrum is significant.

The bottom right panel of Fig.~\ref{fig_matched_spec} shows a close look at the \heii \ line. As seen in the figure, BPASS models including VMS from \cite{fabrice2022} do a better job in reproducing the strength of \heii \ in our sources than models without VMS. However, the \heii \ emission appears much narrower in the models than in our sources. The right panel of Fig.~\ref{fig_matched_spec} shows the \heii \ line widths (FWHM) obtained for our sources which range from $\approx 1000 - 3000$ km s$^{-1}$ with a mean value of $\rm FWHM \simeq 2300$ km s$^{-1}$. These are much larger on average than the line widths obtained from the models which range from $\approx 1000 - 1500$ km s$^{-1}$. The shape of \Heiiuv \ profile of the sources is not fully reproduced by theoretical VMS models from \cite{fabrice2022}; while the shape in \Civuv \ profile matches well. This may have to do with details of the wind properties such as clumping or the exact shape of the velocity structure. This issue should be further addressed in future studies of VMS and related populations.

\section{Discussion}
\label{s_discussion}


\subsection{Incidence of VMS in UV-bright galaxies and in other galaxy populations}

\begin{figure*}
    \centering
    \includegraphics[width=\linewidth]{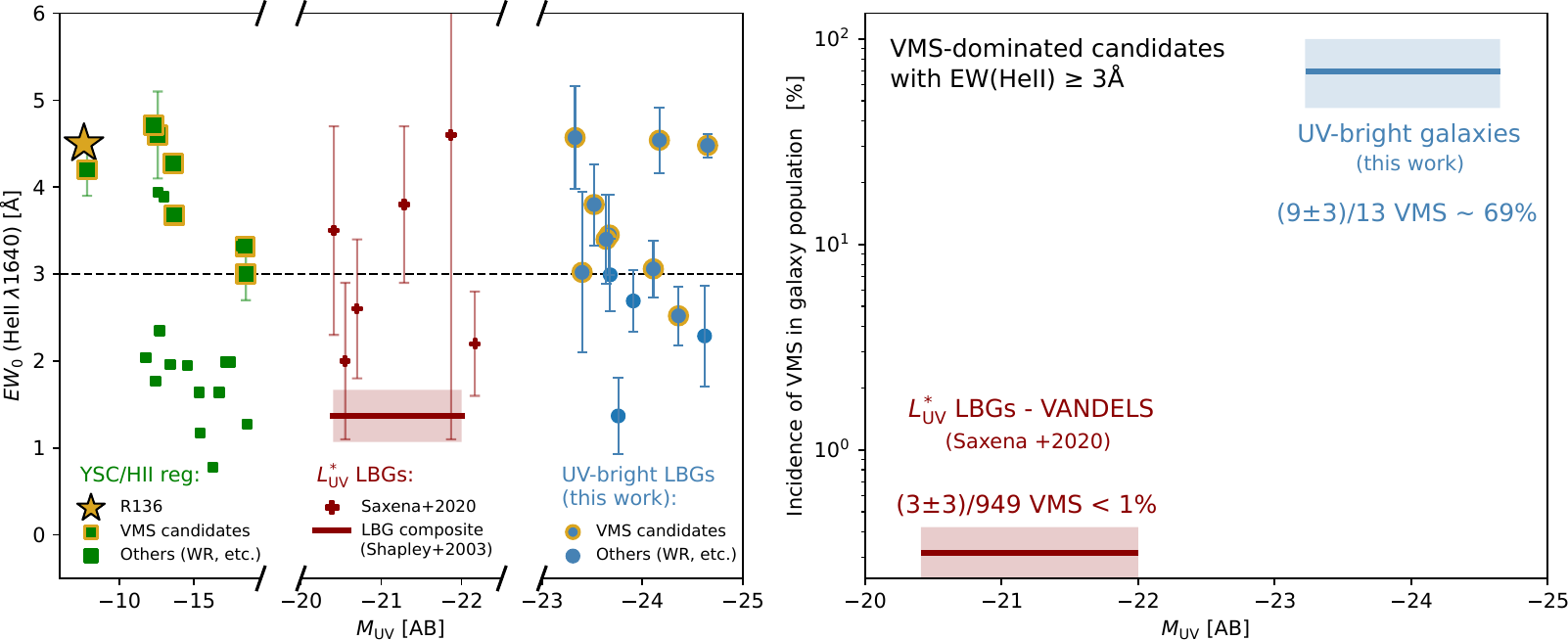}
    \caption{Left: Intensity of the He~{\sc ii} $\lambda 1640$ line for different types of sources: young star clusters and H~{\sc ii} regions (green), normal or typical ($L_{\rm UV}^{*}$) Lyman break galaxies (red), and the UV-bright galaxies studied in this work (blue). The R136 cluster and other VMS-source candidates are marked in yellow. Right: incidence of VMS in different galaxy populations: normal LBGs in red and UV-bright galaxies in blue. 
    }
    \label{fig_last_vms_in_gal_pop}
\end{figure*}

We now investigate the incidence of VMS in different galaxy populations. The left panel of Fig.~\ref{fig_last_vms_in_gal_pop} shows the intensity of the He~{\sc ii} $\lambda 1640$ line for different types of sources. These include young star clusters and H~{\sc ii} regions (green) from the compilation by \cite{Martins2023}, for which we only show VMS- and WR-dominated sources from their classification,\footnote{Here we consider VMS candidates based on sources bearing the labels "VMS" or "VMS or WR" in the classification scheme of \cite{Martins2023}, as detailed in their Table 4, thereby indicating at least sources where VMS are suspected.} normal LBGs (red), and the UV-bright galaxies studied in this work (blue), as indicated by their respective UV absolute magnitudes.

As illustrated in the left panel of Fig.~\ref{fig_last_vms_in_gal_pop}, a significant fraction of UV-bright galaxies exhibit prominent He~{\sc ii} $\lambda 1640$  line with $EW_{0} \simeq 3.0-5.0$ \AA. Similar strengths are also observed in the spectra of local star clusters where VMS are suspected but are significantly higher than those found in WR-dominated clusters. Furthermore, additional spectral features characteristic of VMS, such as the broad N~{\sc iv} $\lambda 1486$ emission and a significant P-Cygni line profile in N~{\sc iv} $\lambda 1719$ \citep[e.g.,][]{crowther2016, fabrice2022}, are observed in the spectra of several UV-bright galaxies. While acknowledging that follow-up observations of the rest-optical blue and red bumps are necessary to definitively establish their VMS nature, as suggested by \cite{Martins2023}, we classify sources with $EW_{0} \geq 3.0$ \AA \ (including J1220+0842, as discussed in Section \ref{ss_empirical_results} and \ref{ss_models_results}) as candidates to host a significant number of VMS. 

It is important to emphasize that our selection of UV-bright sources for GTC follow-up was not based on the presence of strong He~{\sc ii} emission lines. Indeed, He~{\sc ii} $\lambda 1640$ is not detected or barely detected only in their SDSS spectra. Furthermore, the average $EW_{0}$~(He~{\sc ii})~$= 3.24 \pm 0.91$ \AA \ found for our GTC sample is also similar to that of the SDSS stack composed of $\simeq 70$ UV-bright galaxies, where our sources were initially selected ($EW_{0}$~(He~{\sc ii})~$\simeq 3.5$ \AA, R. Marques-Chaves in prep.). This means that our GTC subsample is representative of UV-bright galaxies in general, and suggests that the incidence of VMS-dominated sources in UV-bright galaxies is fairly high, around $\sim 60-70 \%$ (right panel of Fig.~\ref{fig_last_vms_in_gal_pop}). 

A different situation is found in normal/typical ($L_{\rm UV}^{*}$) star-forming galaxies. While \Heiiuv \ is recurrently observed in $L_{\rm UV}^{*}$ galaxies, its strength is substantially weaker than the $EW_{0} = 3.0$ \AA \ threshold used to differentiate the VMS or WR contributions. As already discussed in previous sections, the composite spectrum of \cite{shapley2003} from almost 1000 individual spectra of $z\sim 3$ LBGs shows $EW_{0} \simeq 1.4$ \AA, and has been suggested to be due to WR stars \citep{brinchmann2008pettini, Eldridge2012}. In addition, the composite spectrum of $L_{\rm UV}^{*}$ LBGs does not show any hint of broad emission in N~{\sc iv} $\lambda 1486$ nor N~{\sc iv} $\lambda 1719$ P-Cygni line profiles. This suggests that the average contribution of VMS in $L_{\rm UV}^{*}$ LBGs is likely marginal. These findings are corroborated by the works of \cite{cassata2013}, \cite{Nanayakkara2019}, and \cite{saxena2020}, which provide measurements of $EW_{0}$~(He~{\sc ii}) for individual $z \sim 2-5$ $L_{\rm UV}^{*}$ LBGs using ultra-deep spectroscopy. For example, among nearly 950 sources observed within VANDELS, \cite{saxena2020} find intense ($EW_{0} \gtrsim 3.0$ \AA) and broad ($\rm FWHM > 1000$~km s $^{-1}$) emission in the 1640 \AA \ line for only three of them (Fig.~\ref{fig_last_vms_in_gal_pop}, left). This suggests that the relative number of VMS over OB stars in $L_{\rm UV}^{*}$ star-forming galaxies is low or negligible, with a VMS incidence below $<1 \%$ (right panel of Fig.~\ref{fig_last_vms_in_gal_pop}).

Understanding the underlying factors that account for the prevalence of VMS in UV-bright galaxies but their rarity in $L_{\rm UV}^{*}$ LBGs is certainly important, albeit challenging. Various factors, such as differences in metallicity, age, and star-formation histories between these two classes of sources, may potentially influence our results. We note, however, that metallicity estimations for $L_{\rm UV}^{*}$ LBGs are generally similar to those inferred for our sources \citep[e.g.,][]{pettini2001, steidel2014}. Furthermore, LBGs are, by selection, actively star-forming galaxies, although they can be substantially older than our UV-bright galaxies ($\simeq 10$ Myr). Nevertheless, even considering continuous star formation over 100 Myr, synthesis models with VMS predict $EW_{0}$~(He~{\sc ii}) $ \simeq 2.0$ \AA, which still surpasses the $EW_{0} \simeq 1.4$ \AA \ observed in $L_{\rm UV}^{*}$ LBGs. Therefore, a more natural explanation could be due to intrinsic differences in the IMF, which may enhance the formation of VMS in UV-bright galaxies. 

Recent results do indeed suggest that the IMF can grow toward top-heavy with increasing density and decreasing metallicity \citep[e.g.,][]{Marks2012, Haghi2020, Weatherford2021}. While the metallicity argument may be not valid for our UV-bright galaxies, they appear very compact considering their high stellar masses and SFRs. At least two of these sources (J0121+0025 and J1220+0842) have ground-based optical imaging with very good seeing conditions, for which \cite{marques2020b, marques2021} inferred characteristic sizes of $r_{\rm eff} \lessapprox 1$ kpc. This results in very high stellar mass and SFR surface densities of log($\Sigma_{M_{\star}} / M_{\odot}$ pc$^{-2}) > 2.8$ and log($\Sigma_{\rm SFR} / M_{\odot}$ yr$^{-1}$ kpc$^{-2}) > 2.0$, which are substantially higher than those found typically in $L_{\rm UV}^{*}$ LBGs at similar redshifts \citep[e.g.,][]{Shibuya2015}.

\subsection{Possible caveats}

Since standard methods/calibrations used to derive physical parameters do not include the effects of VMS one may wonder how accurate, for example, the SFR values listed here are. Comparing our models with/without VMS, we find that SFR(UV) could be reduced by $\sim 10-30$ \% for ages between $\sim 10-100$ Myr and maximum stellar masses up to 400 \msun, since VMS boost the UV luminosity but not by very large factors in the case of extended (constant) star-formation. This uncertainty is less than the age-dependence of the SFR(UV) conversion factor $\kappa_{\rm UV}$ over the same age interval (which is a factor $\sim 1.7$). A more detailed discussion of this and other effects of VMS will be presented elsewhere.

We discuss if chemical enrichment from VMS overestimates the metallicities of our sources. \cite{Higgins2023} and \cite{Vink2023} have recently suggested that VMS could significantly pollute the ISM, also in nitrogen, which is used here to determine the metallicity from the N2 indicator. If correct, this would imply that the true stellar metallicity would be lower than inferred. In this case, the strong \Heiiuv\ emission would probably be even more difficult to explain with normal stellar populations since at least the emission from WR stars is known to diminish with decreasing metallicity. The inference about the exact VMS content would then depend on the metallicity-dependence of VMS, both on their stellar evolution and atmosphere properties, which are essentially unknown, although \cite{Smith2023} suggest that VMS signatures could be similar between LMC metallicity and $\sim 0.15$ solar. If the wind density of VMS decreases with metallicity their \heii\ emission should be weaker, and hence the amount of VMS required to explain the observed emission line would be higher than in the present models. In short, if the metallicity of our sources was significantly lower than that of the LMC, we would probably underestimate the amount of VMS.

Another limitation of the present work is that we have no good measure of the IMF, its slope, and maximum at the high mass end. For simplicity, we have assumed that the classical Salpeter slope extends to higher masses. Constraining independently the slope and upper mass limit appears difficult and is probably a degenerate problem. Also, an exact quantification of these properties will require better models of ``normal'' stellar populations, including a proper description of emission from WR stars, as discussed earlier. In any case,  although the contribution of WR stars to \Heiiuv\ is uncertain, we think that the presence of VMS is quite securely established in the objects studied here.

Clearly, to progress on these issues and more firmly quantify the contribution of VMS in our objects and in general, one needs accurate metallicity measurements, ideally abundances of several species to see if any peculiar abundance patterns are found, deep optical spectra to search for the signatures of WR stars which can be distinguished from VMS \citep[cf.][]{Martins2023}, improved spectral synthesis models of normal stellar populations, and VMS models at different metallicities,
as also noted, for example, by \cite{senchyna2021} and \cite{Smith2023}.

\section{Conclusions}
\label{s_conclusion}
In this work, we have investigated the presence of very massive stars (VMS > $100 M_{\odot}$) in 13 UV-bright star-forming galaxies using deep GTC optical and near-IR spectroscopy. These galaxies, with redshifts between $2.2 \lesssim z \lesssim 3.6$, are among the UV-brightest sources known at high redshift, with UV absolute magnitudes ranging between $-23.30$ to $-24.70$. They also present very large SFRs $\simeq 100-1000$ $M_{\odot}$ yr$^{-1}$ and nebular metallicities of 12+log(O/H) $=8.10-8.50$, with a mean value of $\simeq 8.36$. We have analyzed the integrated rest-frame UV spectra with GTC using empirical templates and population synthesis models with and without VMS. From the analysis of these data, we obtained the following results: 

\begin{itemize}

\item The very high S/N rest-frame UV spectra reveal intense and broad \Heiiuv \ emission for all sources, with $EW_{0}$ (\heii) between $\simeq 1.40$ \AA \ and $\simeq 4.60$ \AA \ and line widths of $\simeq 1200-3200$ km s$^{-1}$ (FWHM). These sources exhibit diverse He~{\sc ii} spectral profiles, from nearly Gaussian/symmetric profiles to asymmetric profiles, or more complex ones with multiple emission/absorption peaks. We find a tentative correlation between (O/H) and $EW_{0}$ (\heii) so that stronger \Heiiuv \ emission is found predominantly at higher metallicities. The rest-frame UV spectra also show other spectral features originating from very young stellar populations, such as strong P-Cygni line profiles in the wind lines N~{\sc v} $\lambda 1240$, Si~{\sc iv} $\lambda\lambda 1393,1401$ and C~{\sc iv} $\lambda\lambda 1548,1550$, indicating very young ages on the order of $\sim$10Myr assuming continuous star-formation histories.  

\

\item We compare the GTC spectra and the \Heiiuv \ profiles of our UV-bright galaxies with those of known VMS-dominated sources and other empirical spectra of typical galaxies with normal stellar populations, that is, without VMS. We find that the rest-UV spectra of some of the UV-bright galaxies closely resemble those of VMS-dominated clusters, like the local R136/LMC and SB179 clusters or the Sunburst cluster at $z=2.37$, including the strength and spectral shape of the \Heiiuv \ line. On the other hand, the spectra of UV-bright galaxies differ significantly from those of typical ($L_{\rm UV}^{*}$) galaxies where only normal massive stars are expected (i.e., $<100M_{\odot}$ including WR stars).
Furthermore, the spectra of UV-bright galaxies also reveal other VMS signatures, such as \Nivuvnear \ emission and \Nivuvfar \ P-Cygni line profiles, providing additional evidence for the presence of VMS in these sources.

\

\item We also compare the strengths of the observed \Heiiuv \ emission and the absorption component of the N~{\sc v} $\lambda 1240$ and C~{\sc iv} $\lambda \lambda$1548,1550 of our sources with those of synthetic population spectra. For that, we use both standard BPASS models with an IMF upper mass cutoff of 100 $M_{\odot}$ and updated models incorporating VMS self-consistently \citep{fabrice2022} with upper mass cutoffs up to 475 $M_{\odot}$. We find that the majority of UV-bright galaxies require a contribution of VMS to explain the observed strengths of \heii \, N~{\sc v}, and C~{\sc iv}. Therefore, our results suggest that UV-bright galaxies have a different IMF with upper mass limits between $M_{\rm up}=175-475 M_{\odot}$, assuming a Salpeter slope. 

\ 

\item Using an empirical threshold of $EW_{0}$~(\Heiiuv) $=3.0$ \AA \ to differentiate VMS or WR contributions, along with the detection of other VMS spectral profiles (\Nivuvnear, and \Nivuvfar), we classify nine out of 13 UV-bright galaxies as VMS-dominated sources. This suggests that the incidence of VMS-dominated sources in the UV-bright galaxy population is high, around $\approx 70\%$, and is much higher than in typical $L_{\rm UV}^{*}$ Lyman break galaxies at similar redshifts where the incidence of VMS appears negligible ($<1\%$). 

\end{itemize}

\begin{acknowledgements}
{We thank the anonymous referee for critical reviews of the manuscript. Based on observations made with the Gran Telescopio Canarias (GTC) installed in the Spanish Observatorio del Roque de los Muchachos of the Instituto de Astrof\'{i}sica de Canarias, in the island of La Palma. We thank the GTC staff for their help with the observations. We thank Dr. Eros Vanzella for providing us with the spectrum of the Sunburst cluster. A.U. is grateful for support from the Warwick Astronomy Prize Studentship Fund. E.R.S. is supported in part by UK STFC grants ST/X001121/1 and ST/T000406/1.
}

\end{acknowledgements}


\bibliographystyle{aa}

\bibliography{references}

\appendix

\section{GTC spectra of individual sources}\label{A1_GTC_spectra}

\begin{figure*}
    \centering
    \includegraphics[width=0.98\linewidth]{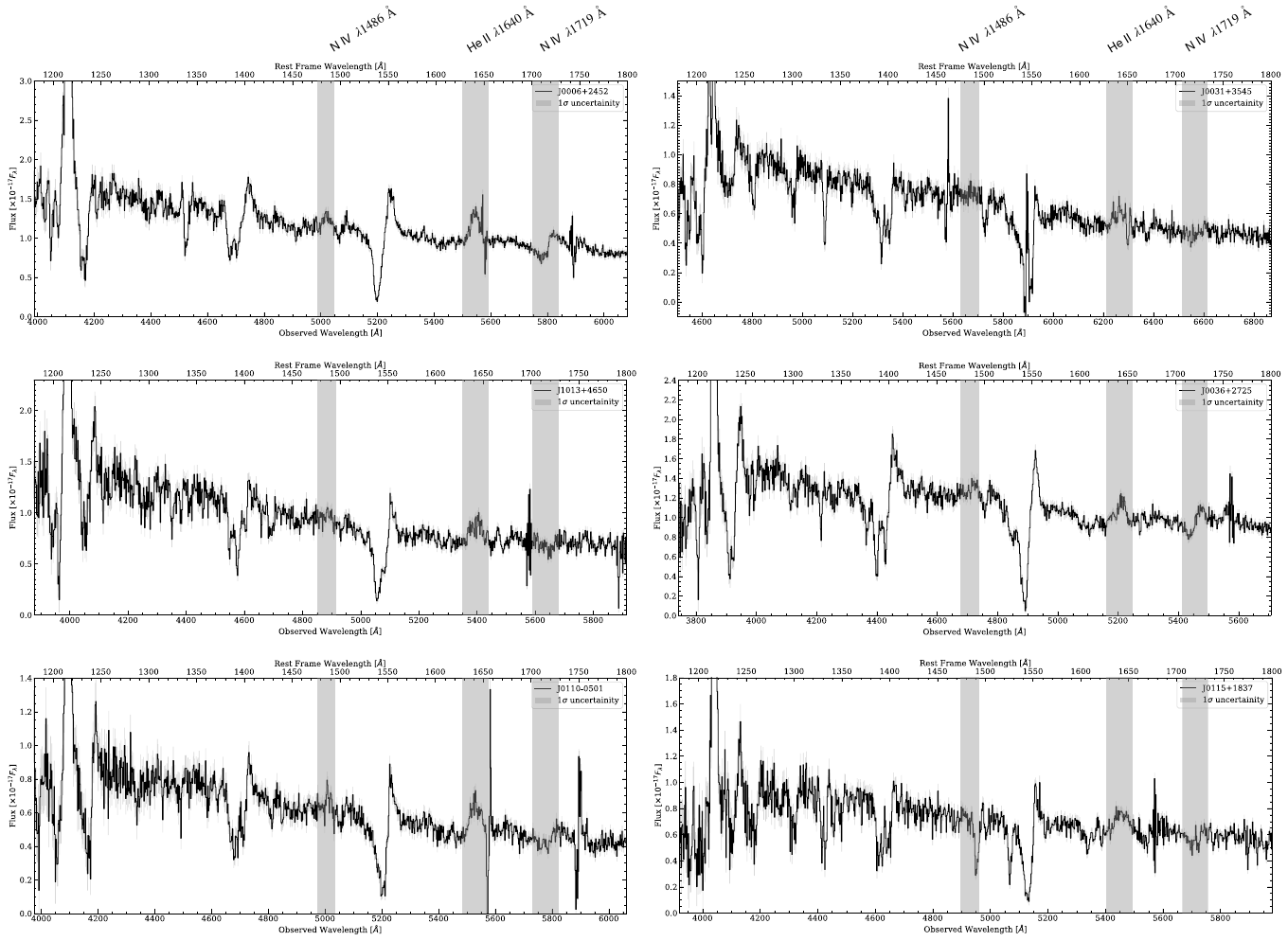}
    \caption{GTC spectra of the six out of 13 UV-bright galaxies studied in this work are in black with the 1$\sigma$ uncertainty in gray. The calibrated spectra have a resolution of around, $\rm R\sim 700$. The x-axes represent the rest frame and observed wavelengths (bottom and top axis, respectively). The three signature VMS profiles, \Heiiuv \ emission, \Nivuvnear \ emission, and \Nivuvfar \ P-Cygni line profile are highlighted in gray shaded regions.}
    \label{spectra_source}
\end{figure*}

\begin{figure*}
    \centering
    \includegraphics[width=0.98\linewidth]{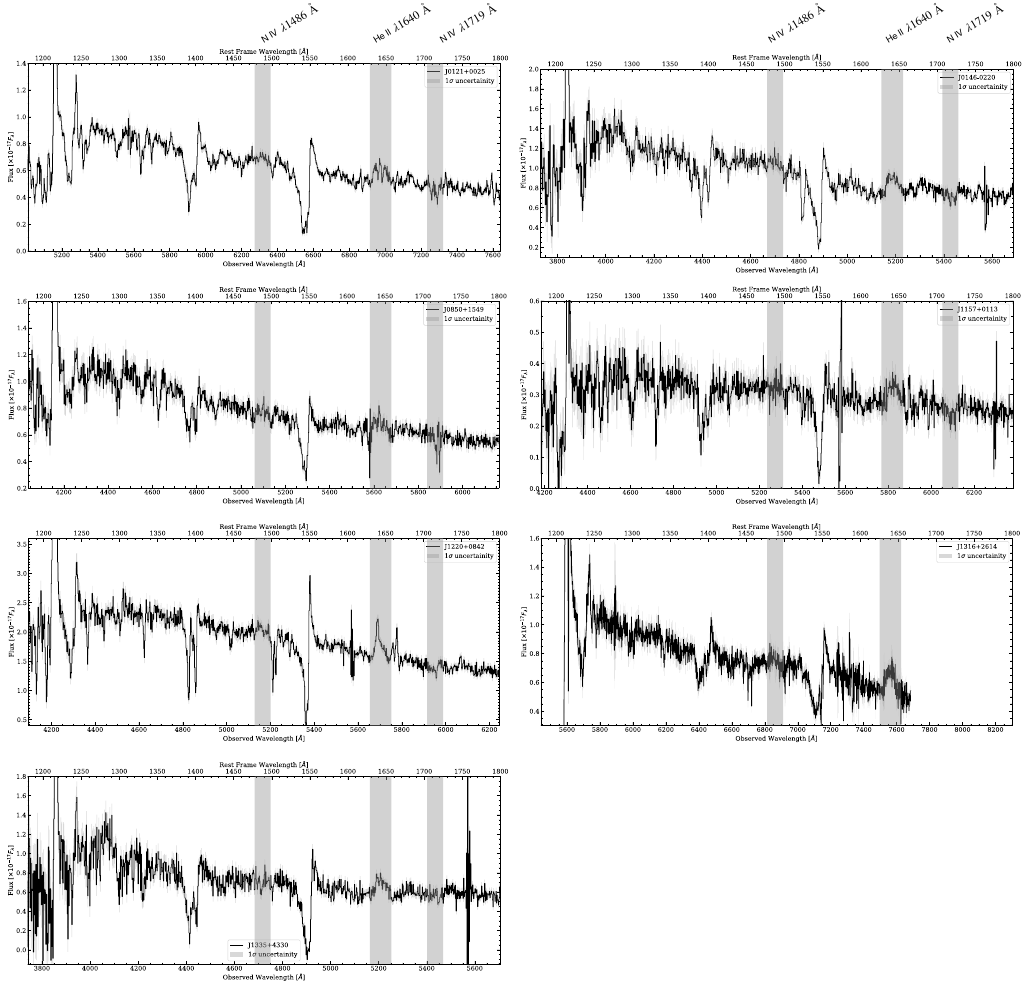}
    \caption{GTC spectra of the seven out of 13 UV-bright galaxies studied in this work are in black with the 1$\sigma$ uncertainty in gray. The calibrated spectra have a resolution of around, $\rm R\sim 700$. The x-axes represent the rest frame and observed wavelengths (bottom and top axis, respectively). The three signature VMS profiles, \Heiiuv \ emission, \Nivuvnear \ emission, and \Nivuvfar \ P-Cygni line profiles are highlighted in gray shaded regions.}
    \label{spectra_source2}
\end{figure*}

Figure \ref{spectra_source} and \ref{spectra_source2} show the GTC optical spectra of the 13 sources located at redshift (z) $\sim$ 2.2 - 3.6 providing the rest-frame UV coverage.

\section{Population synthesis by extrapolating the IMF}
\label{imf_extrapolation}

\cite{fabrice2022} uses an approach IMFs have been extrapolated to different upper mass limits by adding the SEDs of BPASS 100 $M_\odot$ stars with that of the VMS SEDs from \cite{fabrice2022}. In this approach, the number of VMS is calculated in the following mass bins: [100 $M_\odot$, 175 $M_\odot$], [175 $M_\odot$, 225 $M_\odot$], [225 $M_\odot$, 275 $M_\odot$], [275 $M_\odot$, 300 $M_\odot$]. These bins were chosen to add the SEDs of VMS which are only available for 150, 200, 250, and 300 $M_\odot$ stars. The following equation prescribed in \cite{bpass2018} has been used to calculate the number of stars between the mass bin $0.1 M_{\odot}$ and a maximum mass $M_2$,

$$N (0.1, M_2) = C \times \left( \int_{0.1}^{M_1} M^{\alpha_1} \ \mathrm{d}M + M_1^{\alpha_1}\int_{M_1}^{M_2} M^{\alpha_2} \ \mathrm{d}M \ \right).$$

\begin{figure}
    \centering
    \includegraphics[width=0.98\linewidth]{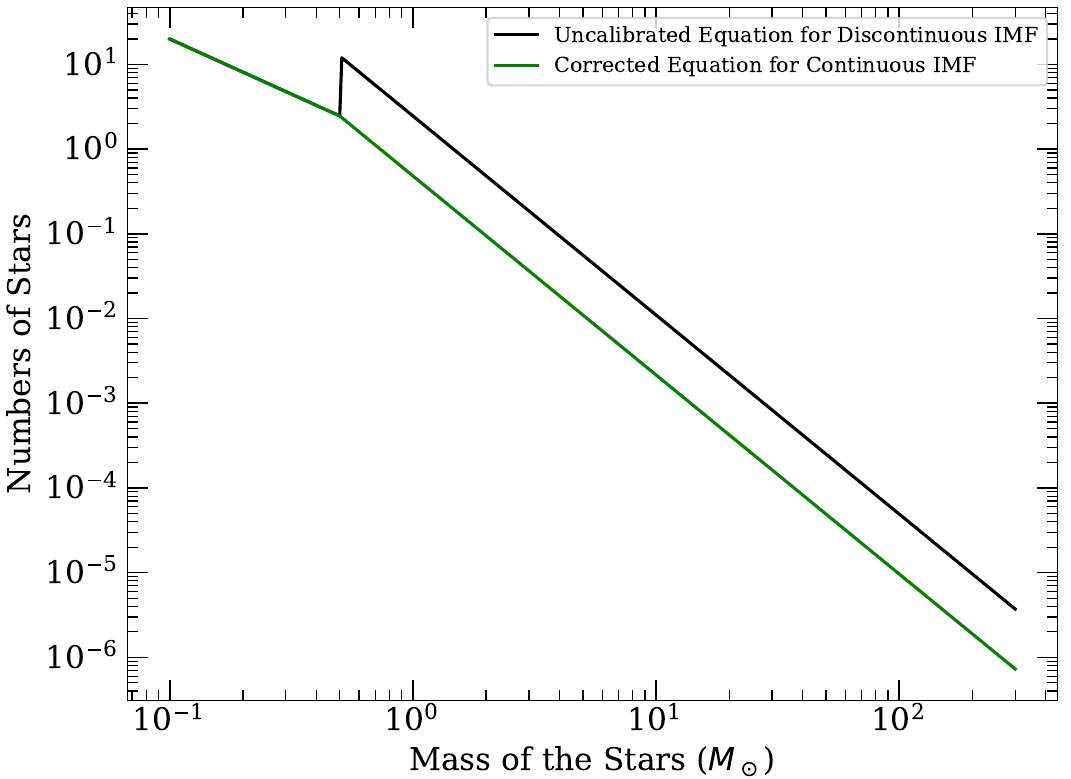}
    \caption{Visual illustration to show the discontinuity in the IMF when using the uncalibrated equation for the Salpeter IMF used in \cite{fabrice2022} and taken from \cite{bpass2018} (in black). The discontinuity goes away when the equation is calibrated for the Salpeter IMF (in green) which is the form used in this work.}
    \label{IMF_fig}
\end{figure}

Upon closer inspection, we noticed that the equation had been calibrated only for a Chabrier IMF. In the case of Salpeter IMF, there is a discontinuity in the IMF slope transition at 0.5 $M_\odot$ which has been shown in Figure \ref{IMF_fig}. The discontinuity arises because of the change in slope of the IMF while transitioning from the lower mass range in 0.1 to 0.5 $M_\odot$ to a higher mass range in 0.5 $M_\odot$ to $M_2$ regime. To produce a continuous IMF in the Salpeter form, the equation should take the following form:

$$N (0.1, M_2) = C \times \left( \int_{0.1}^{M_1} M^{\alpha_1} \ \mathrm{d}M + M_1^{(\alpha_1 - \alpha_2)}\int_{M_1}^{M_2} M^{\alpha_2} \ \mathrm{d}M \ \right).$$

The total mass in the mass bin ($0.1, M_2$) is given by

$$M(0.1, M_2) = C \times \ \left(\int_{0.1}^{M_1} M^{(1+\alpha_1)} \ \mathrm{d}M + M_1^{(\alpha_1 - \alpha_2)}\int_{M_1}^{M_2} M^{(1+\alpha_2)} \ \mathrm{d}M \ \right)$$

$$M(0.1, M_2) = C \times \ \left( \dfrac{M_1^{(2+\alpha_2)}}{(2+\alpha_2)} - \dfrac{0.1^{(2+\alpha_2)}}{(2+\alpha_2)} + \dfrac{M_1^{\alpha_1 - \alpha_2}}{(2 + \alpha_2)} \left[M_2^{2 + \alpha_2}-M_1^{2 + \alpha_1}\right]\right).$$

We have used the new equation calibrated for Salpeter IMF to calculate the number of VMS in different mass bins. We calculate the normalization term $C$ in each case when the IMF upper mass has been extended to different upper mass limits.

In this work, we have defined the following mass bins: $[100 M_{\odot}, 175 M_{\odot}]$, $[175 M_{\odot}, 225 M_{\odot}]$, $[225 M_{\odot}, 275 M_{\odot}]$, $[275 M_{\odot}, 325 M_{\odot}]$, and $[325 M_{\odot}, 475 M_{\odot}]$ for which we calculate the normalization constant $C$ each time. We then calculate the total mass in these mass bins $[M_a, M_b]$ using the above equation calibrated for Salpeter IMF. The number of stars in the mass bin $[M_a, M_b]$ is given by the following equation:

$$N(M_a, M_b) = C \times \left(\dfrac{ \dfrac{M_a^{(\alpha_1 - \alpha_1)}}{2+\alpha_2} \left[M_b^{(2+\alpha_2)} - M_a^{(2+\alpha_2)} \right]}{\dfrac{M_a + M_b}{2}}\right).$$

This gives us 153.52 number of 150 $M_{\odot}$ VMS in the mass bin $[100 M_{\odot}, 175 M_{\odot}]$, 44.41 number of 200 $M_{\odot}$ VMS in the mass bin $[175 M_{\odot}, 225 M_{\odot}]$, 26.04 number of 250 $M_{\odot}$ VMS in the mass bin $[225 M_{\odot}, 275 M_{\odot}]$, 16.85 number of 300 $M_{\odot}$ VMS in the mass bin $[275 M_{\odot}, 325 M_{\odot}]$, and 34.45 number of 400 $M_{\odot}$ VMS in the mass bin $[325 M_{\odot}, 475 M_{\odot}]$.

We note that, the SEDs of stars up to $100 M_{\odot}$ from BPASS are calibrated for a total stellar mass of $10^6 M_{\odot}$. We add these SEDs to the SEDs of VMS from \cite{fabrice2022} to get the new VMS models. This implies in our new VMS models, the luminosity contribution from stars up to $100 M_{\odot}$ is slightly overestimated. In our new VMS models, the VMS contributes around 2.3$\%$ to 5.7$\%$ of the total mass of $10^6 M_{\odot}$. To incorporate this change in mass to SEDs, the SEDs of lower mass stars (M<$100 M_{\odot}$) need to be calibrated by the same percentage of change in mass. As most of these masses will be distributed to the low mass end of the lower mass stars (M<$100 M_{\odot}$) because of the shape of the IMF, we expect little change in overall SED contributions. In any case, the next-generation population synthesis models that will use the \cite{grafener2021} mass loss recipe for the VMS to produce SEDs by incorporating all the mass ranges in their evolution, would be able to solve this discrepancy.

\end{document}